\documentclass[twocolumn,pra,aps,longbibliography]{revtex4-1}

\usepackage{graphicx}
\usepackage{amsmath}
\usepackage{amssymb}
\usepackage{amsfonts}
\usepackage{dsfont}
\usepackage{dcolumn}
\usepackage{bbm}
\usepackage{float}
\usepackage{subfigure}
\usepackage{mathtools}
\usepackage{braket}
\usepackage{placeins}
\usepackage{physics}
\usepackage{longtable}
\usepackage{siunitx}
\usepackage[unicode]{hyperref}
\usepackage[nameinlink]{cleveref}
\Crefname{section}{Sec.}{Secs.}
\Crefname{equation}{Eq.}{Eqs.}
\Crefname{figure}{Fig.}{Figs.}
\Crefname{tabular}{Tab.}{Tabs.}

\usepackage{bbold} 
\usepackage{nicefrac}
\usepackage{tikz}

\newcommand{\matrixsymbol}[1]{#1} 
\newcommand{\Dim}[0]{d} 

\newcommand{\bes}{\begin{subequations}}
\newcommand{\ees}{\end{subequations}}

\newcommand{\erz}[1]{f_{#1}^\dagger}
 
\newcommand{\ver}[1]{f_{#1}^{\vphantom{\dagger}}}

\newcommand{\bzo}[1]{\widehat{n}_{#1}}

\begin{document}

\title{Probing thermalization in quenched (non-)integrable Fermi-Hubbard models}

\author{Philip Bleicker}
\email{philip.bleicker@tu-dortmund.de}
\affiliation{Lehrstuhl f\"{u}r Theoretische Physik I, 
Technische Universit\"{a}t Dortmund,
 Otto-Hahn-Stra\ss{}e 4, 44221 Dortmund, Germany}

\author{Joachim Stolze}
\email{joachim.stolze@tu-dortmund.de}
\affiliation{Lehrstuhl f\"{u}r Theoretische Physik I, 
Technische Universit\"{a}t Dortmund,
 Otto-Hahn-Stra\ss{}e 4, 44221 Dortmund, Germany}

\author{G\"otz S.\ Uhrig}
\email{goetz.uhrig@tu-dortmund.de}
\affiliation{Lehrstuhl f\"{u}r Theoretische Physik I, 
Technische Universit\"{a}t Dortmund,
 Otto-Hahn-Stra\ss{}e 4, 44221 Dortmund, Germany}

\date{\textrm{\today}}

\begin{abstract} 
Using numerically exact methods we examine the Fermi-Hubbard model on arbitrary cluster 
topology. We focus on the question which  systems eventually equilibrate 
or even thermalize after an interaction quench
when initially prepared in a state highly entangled between system and bath. 
We find that constants of motion in integrable clusters prevent equilibration to the 
thermal state. We discuss the size of fluctuations during equilibration
and thermalization and the influence of integrability. The influence of real space topology
and in particular of infinite-range graphs on equilibration and thermalization is studied.
\end{abstract}

\pacs{05.70.Ln, 67.85.−d, 71.10.Fd, 71.10.Pm}

\maketitle

\section{Introduction}
\label{s:introduction}

Non-equilibrium quantum physics is attracting much interest currently. 
This is partly due to the significantly increased experimental possibilities, for
instance in artificial systems of atoms in optical lattices \cite{Anderson1998,Bloch2005,Trotzky2011} 
and by ultrafast pump and probe spectroscopy of condensed matter systems \cite{Axt2004,Morawetz2004,Perfetti2006}. 
Partly, fundamental conceptual issues \cite{Eisert2015} attract theoretical interest more and more: 
How does equilibration and thermalization occur in 
closed quantum systems? Which features of the systems and of the quenches influence 
these processes? Under which circumstances could thermalization even break down 
\cite{Imbrie2017,Abanin2019} or be weakened?
The main aim of the present article is to contribute to
the understanding of these issues by a comprehensive numerical study on clusters.

A common technique to create a non-equilibrium situation is a quench 
\cite{Lauchli2008,Chen2011,Langen2013}, i.e., an abrupt change
of parameters of the system. The main aim of our work is to examine both equilibration 
and thermalization after such quenches in the 
context of the Fermi-Hubbard model with arbitrary topology and to study the prerequisites 
under which the different phenomena occur. Throughout this work, the terms 
\textit{equilibration} and \textit{thermalization} appear. At first glance, they seem to 
refer to the same phenomenon. Thus,  it is worthwhile to point out what is actually meant 
by these terms and in how far they differ from each other.

\subsection{Equilibration}

By equilibration of a quantum system we denote the process that time-dependent observables 
$\expval{A(t)}$ eventually relax for $t\to\infty$ 
to an average value $\overline{A}=\Tr(A\omega)$ where 
$\omega:=\overline{\rho(t)}$ denotes the density operator of the system
averaged over very long time intervals. 
Equilibration is considered a generic phenomenon 
in quantum systems \cite{Reimann2008,Short2010,Wilming2019,Ptaszynski2019}. 
But finite quantum systems with a finite-dimensional Hilbert space are special in 
a rigorous sense because they display a discrete and finite set of eigenvalues. Hence,
the temporal evolution of an arbitrary quantum state and thereby of its expectation
values is governed by frequencies corresponding to these eigenvalues or more precisely to the differences between these eigenvalues. For a finite set of eigenvalues one has a finite set of
possible frequencies so that an oscillatory evolution is induced (except if one starts
by accident from an eigenstate). Rigorously, no  equilibration towards $\overline{A}$
can occur which seems to
indicate that only infinite systems can display equilibration. While  this conclusion
is correct in the strict sense, it does not reflect the range of observable phenomena. Due to the
exponential increase of the dimensionality of the Hilbert space 
with system size already finite, not too large systems 
reflect the behavior of their infinite counterparts.
But there are fluctuations around the long-time averages $\overline{A}$ and their
size and its dependence on the system size constitute an important issue which we will
address below.

For studying equilibration one conventionally starts by 
partitioning a given closed quantum system into a small subsystem and a considerably larger bath
\begin{equation}
	\mathcal{H} = \mathcal{H}_S \otimes \mathcal{H}_B~\text{with}~\Dim_S \ll \Dim_B
\end{equation}
where $\Dim_i:=\dim{\mathcal{H}_i}$. In line with most of the present literature
we assume that this partitioning is done in real space. Certain aspects may 
carry over to other representations as well. Measurements are supposed to take place
on the smaller subsystem which can be taken as small as a single site if
the measurement of local, on-site observables is considered.
Equilibration means that the chosen subsystem $S$ resides in a state described
by the partial density matrix $\rho_S(t)=\Tr_B \rho(t)$ 
which is close to its time-averaged state $\omega_S=\overline{\rho_S(t)}$
for all times, at least after a sufficiently long initial period of relaxation.

For initial product states, i.e., states of the form 
$\ket{\psi}_{SB}=\ket{\psi}_S\otimes\ket{\psi}_B$,
it has been proven rigorously by Linden et al.\ 
\cite{Linden2009} that the trace distance for two Hermitian operators as given by
\begin{equation}
	D(t)=\frac{1}{2}\Tr\left(\sqrt{(\rho_S(t)-\omega_S)^2}\right)
\end{equation}
between 
$\rho_S(t)$ and $\omega_S$ is bounded by
\begin{equation}
	\label{eq:upper_bound_state_distrance}
	\overline{D(t)} \le 
	\frac{1}{2}\sqrt{\frac{\Dim_S}{d_\mathrm{eff}(\omega_B)}} 
	\le \frac{1}{2}\sqrt{\frac{\Dim_S^2}{d_\mathrm{eff}(\omega)}}.
\end{equation}
Here and in similar studies \cite{Reimann2008,Linden2010,Short2010,Short2012,Wilming2019} 
the relevant quantity has proven to be the effective dimension 
$d_\mathrm{eff}(\omega) := 1/\Tr\left(\omega^2\right)$ of the 
time-averaged state $\omega=\overline{\rho(t)}$. The effective dimension is given by
\begin{equation}
	\label{eq:effective_dimension}
	d_\mathrm{eff}(\omega) = \left(\sum_n (\Tr\left(P_n \rho(0)\right))^2\right)^{-1}
\end{equation}
where we use the projector $P_n$ onto the eigenspace of energy $E_n$ 
and the initial state is given by $\rho(0)$ \cite{Short2010,Linden2010}.
If this effective dimension is sufficiently large 
the above inequality implies that the small subsystem equilibrates
in the sense that the expecation values in the subsystem deviate from their
long-time average little and very rarely. 
It is reasonable to presume that the effective dimension in realistic cases of
interacting Hamiltonians is very large due to exponentially many energy 
eigenstates contributing to quenched states, even if these states display 
only small energy uncertainties \cite{Linden2009,Hinrichsen2011}.

The physical motivation for the phenomenon of equilibration in a subsystem is intuitively
accessible. Equilibration means that information encoded in the initial state of the subsystem is lost. Since rigorously the unitary evolution of
the whole quantum system does not allow for information loss, the loss must occur to 
the bath, i.e., to the remainder of the system. This is favored if the quantum dynamics
allows to reach the whole Hilbert space or a substantial part of it. 
This, in turn, implies a high effective dimension.

But even though highly plausible it still remains unclear whether the assumption of a sufficiently 
large effective dimension holds for all physically realistic configurations. What is more, evaluation 
of the quantity $d_\mathrm{eff}(\omega)$ requires an \textit{a priori} 
complete exact diagonalization and thus highly limits the applicability of 
the bound \eqref{eq:upper_bound_state_distrance} in practical situations.

Recent research has reformulated the effective dimension in terms of the R\'enyi entanglement entropy.
This reformulation does not imply an improved calculability \cite{Wilming2019}. In addition, however, an
upper bound for the R\'enyi entanglement entropy was derived which predicts a linear
increase of the entropy with system size $N$, implying an exponential growth
of the effective dimension with $N$. The prefactor of $N$ in these estimates remains yet unknown.

Furthermore, the mathematical considerations implying $\ln(d_\text{eff}) \propto N$
\cite{Wilming2019} consider product states of system and bath
 as initial states. There remains the open issue whether the situation changes fundamentally 
if the system is quenched starting from other types of initial states. 
For this reason, we will investigate 
another generic, but non-product initial state. We prepare the system initially
in a state which is highly entangled with respect to the chosen real space partitioning, 
namely the Fermi sea \eqref{eq:fermi_sea_state}. This means that there are no pure states
of both subsystems $S$ and $B$ individually since $\ket{\text{FS}}$ cannot be split
into a product state of a state of $S$ and $B$, respectively. 
By doing so we intentionally violate one of the main conditions conventionally assumed to hold 
in the process of equilibration. We want to study whether equilibration still occurs in the chosen
more generic setting. Subsequently, the system will be subjected to a quench 
to drive it out of equilibrium in a well-defined and reproducible manner.

\subsection{Thermalization}

When referring to thermalization a specific form of equilibration is meant. If the average value 
$\overline{A}$ equals the thermal value $A_\mathrm{th}$ which results from 
statistical ensemble theory according to
\begin{equation}
	\label{eq:thermal_expectation_value}
	A_\mathrm{th}=\Tr(A \rho_\text{can})
\end{equation}
thermalization has taken place. 
Here the canonical density matrix at inverse temperature $\beta$ reads
\begin{equation}
\label{eq:canonical}
	\rho_\text{can} = \frac{1}{Z}e^{-\beta H}.
\end{equation}
The most intriguing aspect about thermalizing systems 
is that the expecation values $\overline{A} = A_\mathrm{th}$ 
only depend on an effective inverse temperature $\beta_\mathrm{eff}$ 
resulting from the overall energy $E=\expval{H}$ according to
\begin{equation}
	\label{eq:effective_temperature}
	E = - \dv{\beta}\ln(Z(\beta))\big|_{\beta=\beta_\mathrm{eff}}.
\end{equation}
In other words, it appears that the system has lost its memory about the initial
state at $t=0$ except for its energy content. Of course, this cannot be true if
one had access to all conceivable observables of a system. Then it is easy to see 
that this access provides complete knowledge about the temporal evolution of the initial 
state without any loss of information. Hence, equilibration and thermalization can
only occur for observables measured on a small subsystem of the total quantum system.
Typically, observables acting only on a very few adjacent sites are considered. 

Conserved quantities $C_i$ in integrable models restrict the dynamics similar to 
the energy in the canonical ensemble. Obviously, the expectation values of the $C_i$
are constant in time and do not change from their initial values. Thus, they cannot
relax to any thermal value. Hence, thermalization is claimed to be a specific property
of non-integrable systems \cite{Kollath2007,Rigol2008,Rigol2009a}. As an example for the 
constraints of the dynamics of an integrable system we 
consider the Fermi-Hubbard model on a finite chain
 with periodic boundary conditions \cite{Lieb1968,Essler2005} and nearest-neighbor hoppings $J_{ij}=J$
and on-site Hubbard repulsions $U_i=U$. Most of the integrals of motion, but not all of them,
 are functionally dependent on the ratio $\nicefrac{J}{U}$ \cite{Shastry1986,Grosse1989}.
As realizations of non-integrable systems we consider connected clusters of arbitrary topology.

Independent of integrability, any number of integrals of motion $C_i$ restricts 
equilibration. Instead of the thermal density matrix in the canonical ensemble
it is straightforward to derive that the maximization of the entropy of a density matrix
for given, fixed values $\langle C_i\rangle$ for  $C_i$ leads to a generalization
of \eqref{eq:canonical} called the generalized Gibbs ensemble (GGE) \cite{Rigol2007,Rigol2008}
\begin{equation}
	\label{eq:rho_gge}
	\rho_\text{GGE} = \frac{1}{Z}e^{-\sum_i \lambda_i C_i},
\end{equation}
where the $\lambda_i$ are Lagrange multipliers which are determined by the
condition 
\begin{equation}
	\label{eq:condition_gge}
\langle C_i\rangle_\text{GGE} = \langle C_i\rangle_\text{initial}.
\end{equation}
We emphasize that this result does not require that the conserved quantities 
commute pairwise, i.e., $[C_i,C_j]=0$ is not a necessary condition. 
This is so because the entropy to be maximized is given by a trace which allows for cyclic
permutations after derivation so that the sequence of operators can always be chosen
such that $C_i$ stands in front of (or after) the density matrix.
In literature, the GGE for non-commuting integrals of motion is sometimes
called non-Abelian thermal state \cite{YungerHalpern2016,Halpern2019}.
In any case, a system with conserved quantities may show generalized thermalization
to the GGE in \eqref{eq:rho_gge} while its thermalization to the canonical
ensemble \eqref{eq:canonical} is only possible if  this ensemble fulfills
the conditions \eqref{eq:condition_gge} accidentally.

For the scope of the present paper it is of importance to stress the following key idea here once
 again: Non-integrable generic clusters, which are not restricted by any conserved quantities 
other than the overall energy, are expected to show signs of thermalization while integrable ones 
do not. Using numerically exact methods we investigate this expectation for the one-band 
Fermi-Hubbard model in the  remainder of this work. 

This article is structured in the following way: 
In \Cref{s:model}, the Fermi-Hubbard model and the general quench protocol are 
explained briefly. \Cref{s:method} outlines the concepts and algorithms
used to access the time-evolution of observables and their thermal expectation values, 
namely the Chebyshev expansion technique (CET), the kernel polynomial method (KPM)
 and thermal pure quantum states (TPQ). In \Cref{s:results} we discuss the results 
for the globally quenched Fermi-Hubbard model on clusters of 
various topologies, study the influence
 of the cluster properties on the general relaxation behavior and work out 
the thermalization behaviour of different systems.
Summary and outlook are given in \Cref{s:summary}.

\section{Model} 
\label{s:model}

The Fermi-Hubbard model is one of the paradigmatic models for interacting electrons 
on a lattice and combines tight-binding electrons with a 
strongly screened Coulomb interaction \cite{Hubbard1963,Kanamori1963,Gutzwiller1963}. 
In the following, we restrict considerations to the one-band model on 
arbitrarily shaped clusters such that the Hamiltonian takes the form
\begin{equation}
	\label{eq:fh_model}
	H = H_0 + H_\text{int} = - \sum_{\substack{ij\sigma}} J_{ij}{\vphantom{\dagger}}
	\erz{i\sigma} \ver{j\sigma} + \sum_{i} U_i \bzo{{i\uparrow}}\bzo{{i\downarrow}}.
\end{equation}
Here, $\erz{i\sigma}$ ($\ver{i\sigma}$) are the creation (annihilation) operators
at site $i$ for a fermion of spin $\sigma$ and $\bzo{{i\sigma}}$ is the corresponding
number operator, $J_{ij}$ denotes the real hopping matrix element between the sites $i$ 
and $j$ and $U_i>0$ is the on-site interaction, i.e., the energy cost of a double occupancy
 at site $i$. Since we are dealing with arbitrarily shaped clusters, it is convenient to 
introduce the cluster as the undirected graph $G=(V,E)$ consisting of consecutively 
labeled vertices (sites) $V$ each carrying information about its local repulsion $U_i$ 
and edges (hopping matrix elements) $E$. The most natural representation of an 
undirected graph $G$ is by means of its weighted adjacency matrix $A(G)$. 
Here, the weights carry the information about the different hopping strengths $J_{ij}$.

To excite the system to a non-equilibrium state a sudden global interaction quench is
used for which we initially prepare the system in the Fermi sea state $\ket{\text{FS}}$ 
as an eigenstate of $H_0$ and suddenly turn on the local interaction.
Consequently, the quench protocol reads 
\begin{equation}
	\label{eq:quench}
	H_\text{Q}(t) = H_0 + \Theta(t) H_\text{int}
\end{equation}
where $\Theta(t)$ is the Heaviside function. Since the quench in 
$H_\text{Q}(t)$ changes the overall system parameters $U_i$ and thus influences 
all sites it is called a global quench. Global quenches have been considered 
to a great extent
\cite{Manmana2007,Moeckel2008,Moeckel2009,Manmana2009,Barmettler2009,Calabrese2011,Calabrese2012a,Caux2013}.

As $H_0$ concerns interaction-free particles diagonalizing $H_0$
 is a one-particle problem. It is sufficient to diagonalize 
the one-particle Hamiltonian $h_0:=-A(G)$ 
in order to obtain the Fermi sea. Let $\ket{i\sigma}$ be the eigenstates 
of the number operator of site $i$ and spin $\sigma$ and 
let $h_0$ fulfill the eigenvalue equation 
$h_0 \ket{\nu\sigma} = \epsilon_\nu \ket{\nu\sigma}$
where $\nu$ is labeling the eigenstates.
 Then, the Fermi sea is constructed by gradually filling the states 
$\ket{\nu\sigma}$ in order of increasing eigenenergies $\epsilon_\nu$ according to
\begin{equation}
	\label{eq:fermi_sea_state}
	\ket{\text{FS}} := \prod_{\mathclap{(\nu,\sigma)\,\in\,I}} 
	\erz{\nu\sigma} \ket{0} 
					=  \prod_{\mathclap{(\nu,\sigma)\,\in\,I}}\,\,\,\, 
					\left(\sum_i \braket{i\sigma}{\nu\sigma} \erz{i\sigma}\right) \ket{0}.
\end{equation}
The index set $I$ is chosen such that the condition 
$\epsilon_\nu < \epsilon_\mathrm{F}$ with $\epsilon_\mathrm{F}$ 
being the Fermi energy is fulfilled for all occupied eigenstates of $h_0$. In the case of degeneracy we construct all possible Fermi sea states and use equal weights for them in the initial density matrix $\rho(0)$.

\section{Method} 
\label{s:method}

In this section, we present a brief overview over the methods used to calculate 
the time-dependence and the thermal expectation values of observables as well 
as the predictions of canonical ensemble theory. Importantly, we point out the 
strengths and shortcomings of the techniques used. 
More detailed mathematical derivations can be found in the references given.

\subsection{Chebyshev expansion technique}
\label{ss:cet}

To obtain the time-dependence $O(t)$ of an observable we resort to Chebyshev expansion 
technique \cite{Tal-EzerH.1984}\,(CET) which consists of the expansion of the unitary 
time evolution operator $U=e^{-iHt}$ in terms of Chebyshev polynomials
\begin{subequations}
\label{eq:chebyshev_polynomials_recursion_relation}
\begin{alignat}{3}
	T_0(x) &= 1, \qquad T_1(x)=x \\
	T_{n+1}(x) &= 2x T_n(x)-T_{n-1}(x)
\end{alignat}
\end{subequations}
which are defined on the closed interval $I=\left[-1;1\right]$. To be able to apply this
 technique to a general Hamiltonian $H$ as in \eqref{eq:fh_model} a finite rescaling 
$H\to H' = (H-b)/a$ is a prerequisite. This is to ensure that the Chebyshev polynomials
 can be used as an orthonormal basis set. In order to perform an appropriate rescaling
 an estimate  of the extremal eigenvalues \cite{Lanczos1950,Arnoldi1951,Kuczynski1992} 
of $H$ is needed to obtain $a=\nicefrac{1}{2}\left(E_\mathrm{max}-E_\mathrm{min}\right)$ 
and $b=\nicefrac{1}{2}\left(E_\mathrm{max}+E_\mathrm{min}\right)$.
Note that estimates in form of upper bounds for $E_\mathrm{max}$ and lower bounds
for $E_\mathrm{min}$ are sufficient because the rescaling only has to ensure that
the rescaled eigenvalues all lie within $I$. Finally, the time-evolution operator 
 becomes
\begin{subequations}
\label{eq:cet_time_dependent_series_final}
\begin{alignat}{3}
	U &= \sum_{n=0}^\infty \alpha_n(t)T_n(H') 
	\\
	\alpha_n(t) &= (2-\delta_{n,0}) i^n e^{-i b t} J_n(at) 
	\label{eq:cet_time_dependent_series_final_coefficients}
\end{alignat}
\end{subequations}
where the time-dependent coefficients essentially depend on the
Bessel functions of the first kind $J_n(at)$.
The dynamics of an initial state $\ket{\psi_0}$ is given by
\begin{equation}
	\label{eq:cet_time_evolution_state}
	\ket{\psi(t)}=U\ket{\psi_0}=\sum_{n=0}^\infty \alpha_n(t)
	\underbrace{T_n(H')\ket{\psi_0}}_{=:\,\ket{\phi_n}}
\end{equation}
with the basis states of the expansion $\ket{\phi_0}\!=\!\ket{\psi_0}$ and 
$\ket{\phi_1}=H'\ket{\psi_0}$ as well as $\ket{\phi_{n+1}}=2H'\ket{\phi_n}-\ket{\phi_{n-1}}$.

Numerically, the infinite series must be cut-off at some finite value 
$N_\mathrm{c}<\infty$. The time dependence of the  prefactors 
is essentially determined by the time dependence \cite{Olver2019} of the Bessel functions $J_n(t)$.  
The higher the order $n$ the longer it takes the Bessel function $J_n(t)$ to 
yield a noticeable contribution to the series. Hence, an estimate for the
accuracy of the truncated series with cut-off value of $N_\mathrm{c}$ can be
given
\begin{equation}
	\epsilon \lessapprox \left(\frac{a t \cdot e }{2 N_\mathrm{c}}\right)^{N_\mathrm{c}}.	
\end{equation}
Consequently, the truncation error is not only related to $N_\mathrm{c}$, but depends
also directly on the maximum time up to which results are calculated as well as on the 
parameter $a$ which equals half the width of the energy spectrum.
Importantly, increasing $N_\text{c}$ linearly increases the time $t$ up to which the
error estimate is the same.

\subsection{Kernel polynomial method}
\label{ss:kpm}

The main aim in the application of the kernel polynomial method 
\cite{Weisse2006}\,(KPM) and of thermal pure quantum\,(TPQ) states \cite{Sugiura2013} 
is to obtain thermal expectation values without the necessity to fully diagonalize the
Hamiltonian. A brief comparison of the results of these two approaches
in different temperature 
ranges will be given in the next \Cref{ss:tpq}.

For KPM we resort again to the rescaled Hamiltonian $H'$ as given in \Cref{ss:cet} 
such that all energies $E\in I$. For brevity, we omit the superscripts from now on. 
Given the canonical partition function
\begin{equation}
	\label{eq:kpm_thermal_expectation_value_partition_function}
	Z = \int_{-1}^{1} \rho(E) e^{-\beta E}\dd{E}
\end{equation}
the desired thermal expectation value becomes
\begin{equation}
	\label{eq:kpm_thermal_expectation_value_observable}
	\expval{O}_\mathrm{th} = \frac{1}{Z} \int_{-1}^1 o(E)e^{-\beta E} \dd{E}.
\end{equation}
The problem consists in finding suitable approximations of the (rescaled) 
density of states $\rho(E)$ and of the observable density $o(E)$ given by
\begin{subequations}
\label{eq:kpm_functions_for_expansion}
\begin{alignat}{2}
	\rho(E) &= \frac{1}{\Dim} \sum_{i=0}^{\Dim-1} \delta(E-E_i) \\
	o(E) &= \frac{1}{\Dim}\sum_{i=0}^{\Dim-1}\mel{i}{O}{i}\delta(E-E_i)
\end{alignat}
\end{subequations}
where $\Dim\!:=\!\operatorname{dim}\qty(\mathcal{H})$ denotes the dimension of the
 Hilbert space. To obtain appropriate
approximations, we expand the real functions \eqref{eq:kpm_functions_for_expansion} as
\begin{subequations}
\begin{alignat}{3}
	f(E)  &=\frac{1}{\pi\sqrt{1-E^2}}\left(\mu_0+2\sum_{n=1}^\infty \mu_n T_n(E)\right)
	\label{eq:kpm_expansion_function}
	\\
	\mu_n &= \int_{-1}^1 f(E) T_n(E) \dd{E} 
	\label{eq:kpm_expansion_moments}.
\end{alignat}
\end{subequations}

The most detrimental effect of truncating  infinite series such as 
the one in \eqref{eq:kpm_expansion_function} after $k<\infty$ terms are Gibbs' oscillations. In the vicinity 
of points where the function to be approximated possesses singularities, for instance
discontinuities, the truncated series displays strong oscillations. This leads to 
unsatisfactory approximations of $f(E)$ and may spoil the integration of the
approximated function with high precision which is necessary for the determination of 
thermal quantities. As a remedy, we convolve \eqref{eq:kpm_expansion_function} 
with the Jackson kernel \cite{Jackson1911,Jackson1912} as introduced 
 by Weisse et al.\ \cite{Weisse2006}. This amounts to 
rescaling the Chebyshev moments of the expansion according to 
$\mu_n \to g_n\mu_n$ with $\alpha:=k+1$ by
\begin{equation}
	g_n = \frac{(\alpha-n)\cos\left(\nicefrac{\pi n}{\alpha}\right)+
	\sin\left(\nicefrac{\pi n}{\alpha}\right)+
	\cot\left(\nicefrac{\pi}{\alpha}\right)}{\alpha}. 
\end{equation}
For the calculation of the moments of the expansion
\begin{alignat}{3}
	\mu_n = \int_{-1}^1 o(E) T_n(E) \dd{E} = 
	\frac{1}{\Dim}\Tr\left(O T_n(H)\right) 
	\label{eq:thermal_moment_trace} 
\end{alignat} 
we employ stochastic trace evaluation as initially proposed by 
Skilling \cite{Skilling1988} and later generalized by others 
\cite{Drabold1993,Silver1994}. 
It consists of the approximation of the full trace 
$\Tr(A)$ by $R\ll d$ randomly chosen quantum states, see also
next section.

\subsection{Thermal pure quantum states}
\label{ss:tpq}

This approach relies on what is called  quantum typicality these days.
It is based on two ingredients. The first is actually the stochastic 
evaluation of traces \cite{Skilling1988,Drabold1993,Silver1994} in order to 
compute thermal averages of quantum mechanical observables in the canonical ensemble.
This element was already used in the KPM approach.  
The second lies in the evolution of stochastic states in imaginary time
to determine the thermal pure quantum states (TPQ).

Using a set of $R$ normalized states $\ket{r}$ whose complex coefficients are each 
drawn from a normal distribution we approximate traces by the average of the
 expectation values
\begin{equation}
	\label{eq:tpq_trace_estimate}
	\Tr\left(O\right) = \Dim \overline{\mel{r}{O}{r}}
\end{equation}
where the overbar denotes the process of determining the arithmetic mean from the set 
of all different random states $\left\{\ket{r}\right\}$ and $\Dim$ 
stands for the dimension of the Hilbert space. 

A central idea in TPQ is to decompose the  application of the Boltzmann weight
to the random state $\ket{r}$ into two contributions for bra and ket. 
The invariance of the trace under cyclic permutations ensures that this 
is correct 
\begin{subequations}
\begin{align}
& \Tr(O\exp(-\beta H)) 
\nonumber
\\
&\qquad \qquad =\
\Tr(\exp(-\beta H/2)O\exp(-\beta H/2))
\\
&\qquad \qquad =\
\Dim \overline{\mel{r}{\exp(-\beta H/2)O\exp(-\beta H/2)}{r}} .
\end{align}
\end{subequations}
Defining
\begin{equation}
	\label{eq:tpq_state}
	\ket{\beta} := \exp(-\beta H/2)\ket{r},
\end{equation}
the partition sum $Z$ can be expressed by $\overline{\braket{\beta}{\beta}}$
and the thermal expectation value itself is given by
\begin{equation}
	\label{eq:tpq_observable}
	\expval{O} = \frac{\overline{\mel{\beta}{O}{\beta}}}
	{\overline{\braket{\beta}{\beta}}}.
\end{equation}
The standard deviation of the estimate \eqref{eq:tpq_observable} scales like
 $1/\sqrt{R\Dim}$. 

A crucial advantage of this technique is the possibility to easily evaluate 
the TPQ states $\ket{\beta}$ without fully diagonalizing $H$ first. 
An especially efficient way \cite{Wietek2019} to determine the matrix exponentials
is by resorting to the Lanczos algorithm \cite{Lanczos1950} to 
approximate the Hamiltonian by its matrix form in the 
Krylov space $\mathcal{K}^{s}\!\left(\ket{r}\right)=
\mathrm{span}\left(\ket{r},\,\matrixsymbol{H}\ket{r},
\,\matrixsymbol{H}^2\ket{r},\ldots,\,
\matrixsymbol{H}^{s-1}\ket{r}\right)$
We draw sufficiently many states $\ket{r}$ to gain a deviation below the tolerance
$\num{e-3}$, i.e., achieving $1/\sqrt{R\Dim}\lessapprox \num{e-3}$,
and compute an adequately large Krylov space of dimension $s$ in each step;
the dimension $s$ must be chosen sufficiently large in order to 
ensure that the systematic error due to $s<d$ is less than the required tolerance. 
For best efficiency, the systematic error is chosen of the same
order of magnitude as the stochastic error.

\begin{figure}
  \includegraphics[trim=10 5 50 50,clip,width=\columnwidth]
	{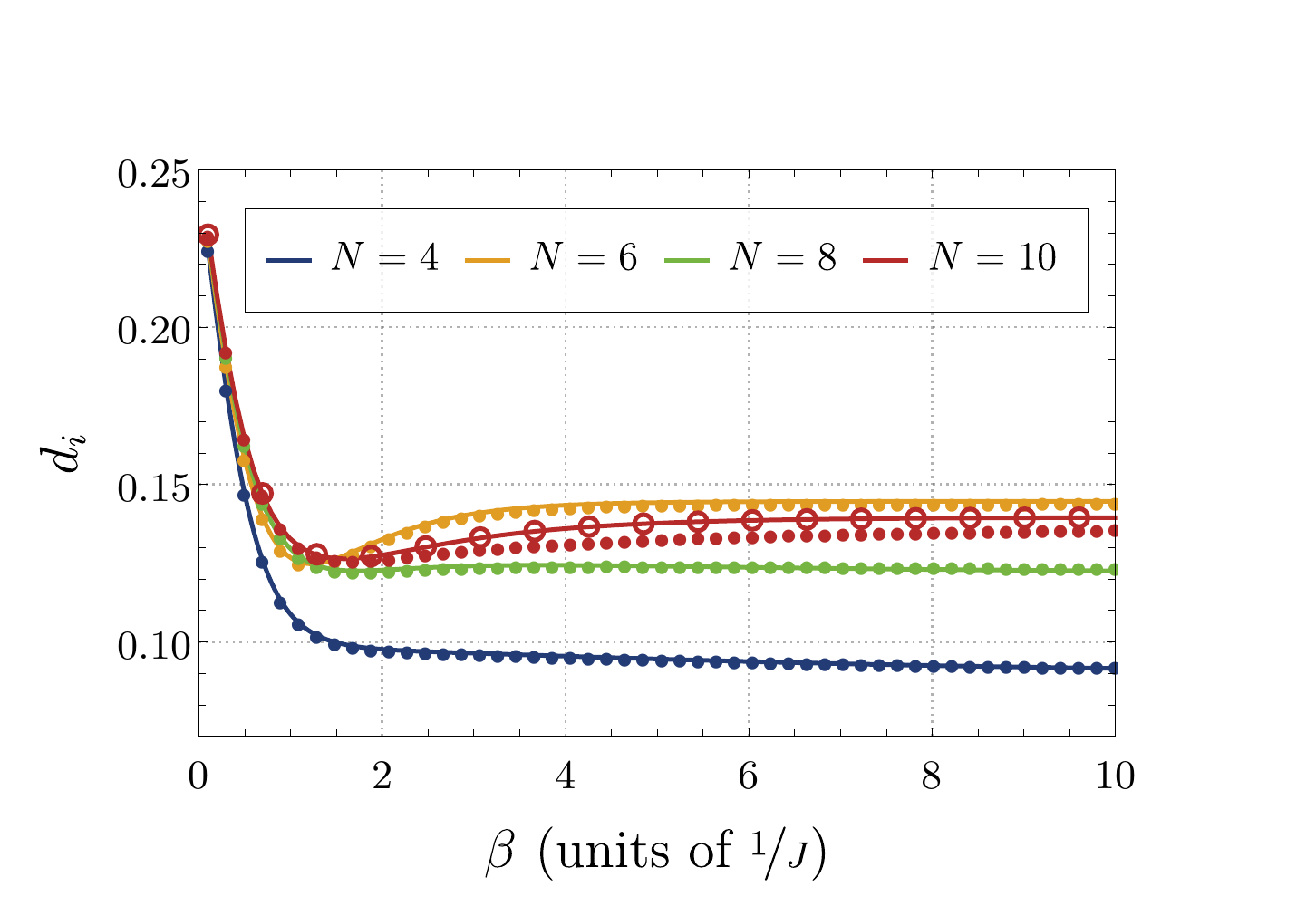}
  \caption{(Color online) Comparison of the results for the thermal expectation value 
	\eqref{eq:thermal_expectation_value} of the double occupancy $d_i$ stemming from 
	exact diagonalization (solid lines) and TPQ states (circles) for a 
	one-dimensional chain with periodic boundary conditions at $U_i=U=3J=3J_{ij}$. 
	All TPQ results are calculated using a Krylov space dimension of 
	$s=10$ except for the open circles in the $N=10$ case ($s=50$).}
  \label{img:tpq_gauge_N_4_to_10_pbc_tpq_vs_ed}
\end{figure}

In \Cref{img:tpq_gauge_N_4_to_10_pbc_tpq_vs_ed} we exemplarily checked the convergence of
TPQ results (circles) against results from exact diagonalization (solid lines) at
half-filling and lattice sizes up to $N=10$. Even for comparably small Krylov dimensions 
$s=10$ a good agreement up to $N=8$, corresponding to $d=4900$ of the
full Hilbert space, can be achieved. Only for $N=10$ sites ($d\approx\num{6.4e4}$) 
the results of TPQ states start to deviate from the exact results. Using a Krylov space 
with $s=50$ is enough as a remedy leading again to a good convergence. 
This observation is in full accordance with the expectation that only small 
fractions of the overall Hilbert space are needed to yield accurate results in 
Krylov space procedures. 

Before comparing  KPM and TPQ to  each other we point out that 
low temperatures $T$ result in large relative Boltzmann weights in 
\Cref{eq:kpm_thermal_expectation_value_partition_function,eq:kpm_thermal_expectation_value_observable} 
for the low-energy part of the spectrum approximated by KPM. 
This, in turn, heavily amplifies even small numerical errors of stochastic or 
systematic origin. This spoils numerical results altogether for low temperatures. 
This issue is fundamental and cannot be solved by trivial means such as increasing
the number of moments $\mu_n$. Approaches have been suggested to overcome 
these obstacles in the interacting regime by combining partial exact diagonalization 
and the kernel polynomial method \cite{Weisse2006}. The ground state and the $m-1$
 energetically lowest excitations of the systems are treated exactly
 while the remainder of the spectrum is calculated using KPM. 

\begin{figure}[htb]
  \includegraphics[trim=10 5 50 50,clip,width=\columnwidth]{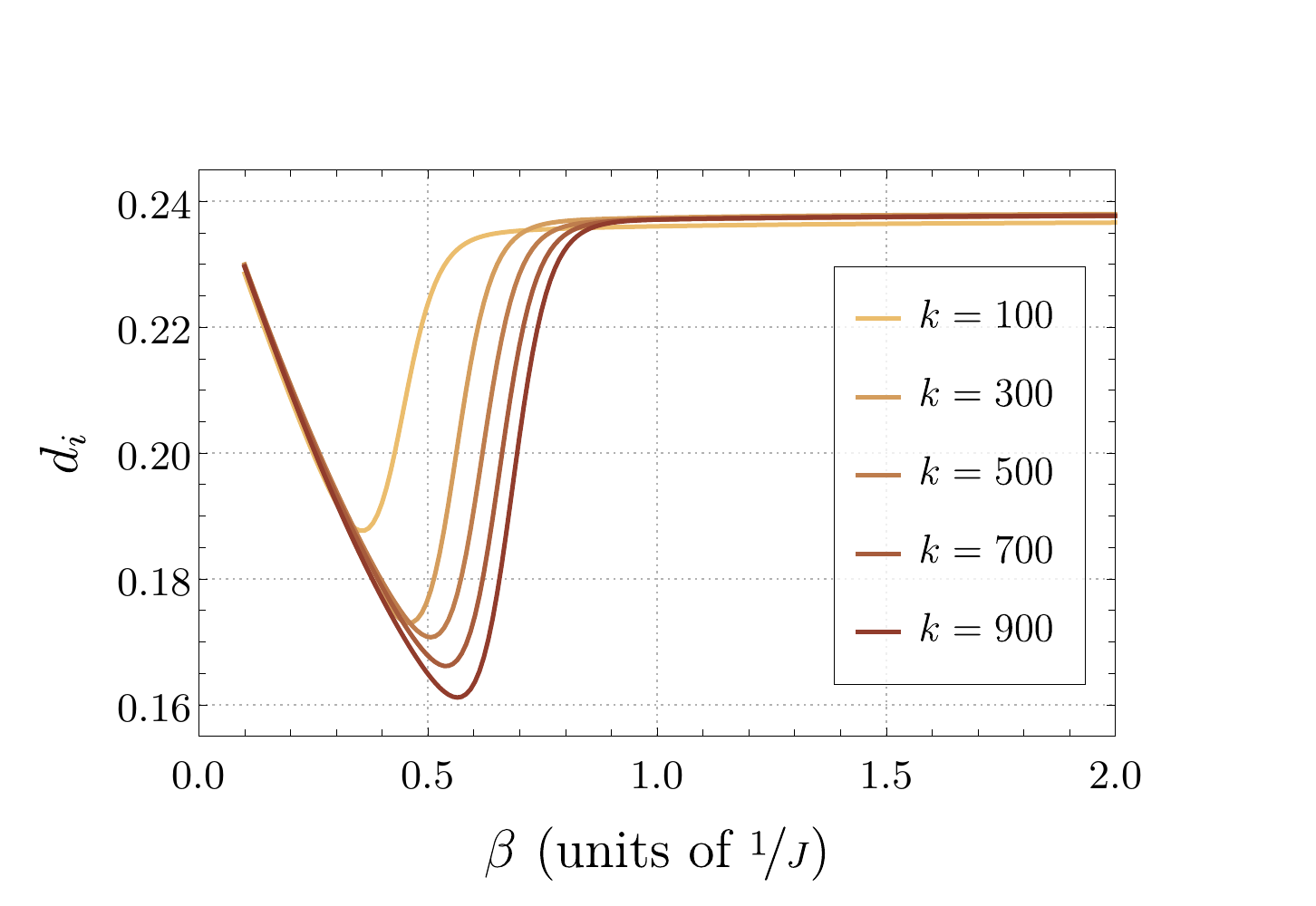}
  \caption{(Color online)\,Thermal expectation value of the double occupancy in a half-filled $N=10$
	Hubbard chain at $U=3J$ calculated by KPM using 
	$k$ moments $\mu_n$. The number of moments increases from top to bottom. A higher number of moments improves the accuracy,
	but it does not change the overall behavior for $\beta J>1$.}
  \label{img:kpm_n10_pbc_hf_U3_different_moments}
\end{figure}

We illustrate the sketched caveat of KPM in its unmodified form described in 
\Cref{ss:kpm}. Specifically, \Cref{img:kpm_n10_pbc_hf_U3_different_moments} displays 
 KPM results which reproduce the physics at lower temperatures (larger
 inverse temperatures $\beta$) slightly better for a higher number $k$ of moments.
But the true low-temperature limit, cf. \Cref{img:tpq_gauge_N_4_to_10_pbc_tpq_vs_ed}, 
is not captured at all. The degree of double occupancy is significantly overestimated
by KPM. 

\begin{figure}[htb]
\includegraphics[trim=10 5 50 50,clip,width=\columnwidth]{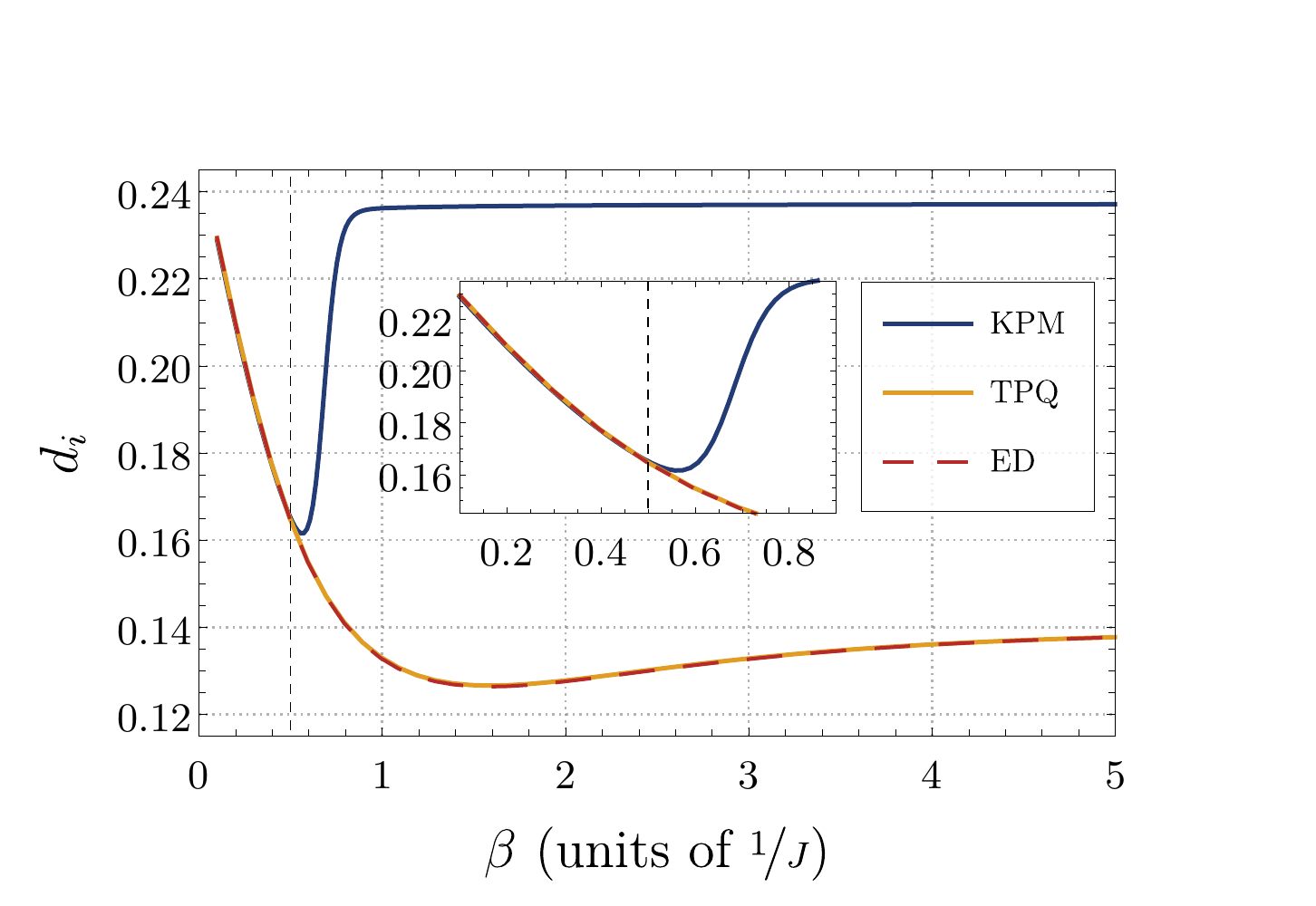}
  \caption{(Color online)\,Comparison of results from  KPM ($k=900$) and TPQ states 
	(Krylov space dimension $s=50$) with results from exact diagonalization (ED). 
	KPM results show notable deviations for increasing $\beta$ while TPQ does not.
	Results for a half-filled Hubbard chain with periodic boundary conditions
	of $N=10$ sites at $U=3J$.}
  \label{img:kpm_vs_tpq_vs_ed_n10_pbc_hf_U3_highest_accuracies}
\end{figure}

Nevertheless, we stress that the high-energy physics is captured very well by KPM. 
To emphasize this point the most accurate results of KPM for $k=900$ and the 
exact results (ED) are compared in the inset of 
\Cref{img:kpm_vs_tpq_vs_ed_n10_pbc_hf_U3_highest_accuracies}. Although KPM results 
start deviating from the ED results at about $\beta\approx0.5$ 
 the high-energy physics for $T\gtrapprox2J$ 
 is described very accurately by KPM.

In contrast to KPM, using TPQ states is very robust against accumulating 
numerical errors since neither a functional approximation based on a truncated series
nor a numerical integration is involved.
In addition, TPQ states are easy to deal with. This makes this approach advantageous. 
Its convergence has been examined in detail \cite{Sugiura2013} 
indicating that the choice of the individual random states $\ket{r}$ 
has an exponentially small effect at finite temperatures
 and that results from TPQ states converge to the actual ensemble 
results exponentially fast in the system size $N$. Thus, TPQ states can be used 
with predictable accuracy leading to well controlled
 results for a broad range of temperature as shown in 
\Cref{img:kpm_vs_tpq_vs_ed_n10_pbc_hf_U3_highest_accuracies}. 
For this reason, all computations of thermal expectation values in the remainder
of this article are performed using TPQ states. 
Two sources of errors need to be controlled: (i) the stochastic error in 
the evaluation of the traces and (ii) the systematic error in the 
evaluation of the matrix exponentials in Krylov spaces of finite dimension $s$.

\section{Results}
\label{s:results}


In this section, we tackle the issues of equilibration and thermalization
on finite clusters. Temporal averages of expectation values and the corresponding
temporal fluctuations will be discussed. The first section deals with equilibration,
the subsequent one with thermalization.

\subsection{Equilibration}
\label{ss:equilibration}

As outlined in the Introduction, analytic arguments for equilibration have been brought forward for the case of initial product states of system and bath \cite{Wilming2019}.
In order to extend evidence for equilibration beyond this special situation,
 we focus on the Fermi sea $\ket{\text{FS}}$
as generic non-product state in real space representation.
The initial non-equilibrium is generated by an interaction quench according to 
\eqref{eq:quench}.

\begin{figure}[htb]
  \includegraphics[trim=10 7 50 50,clip,width=\columnwidth]{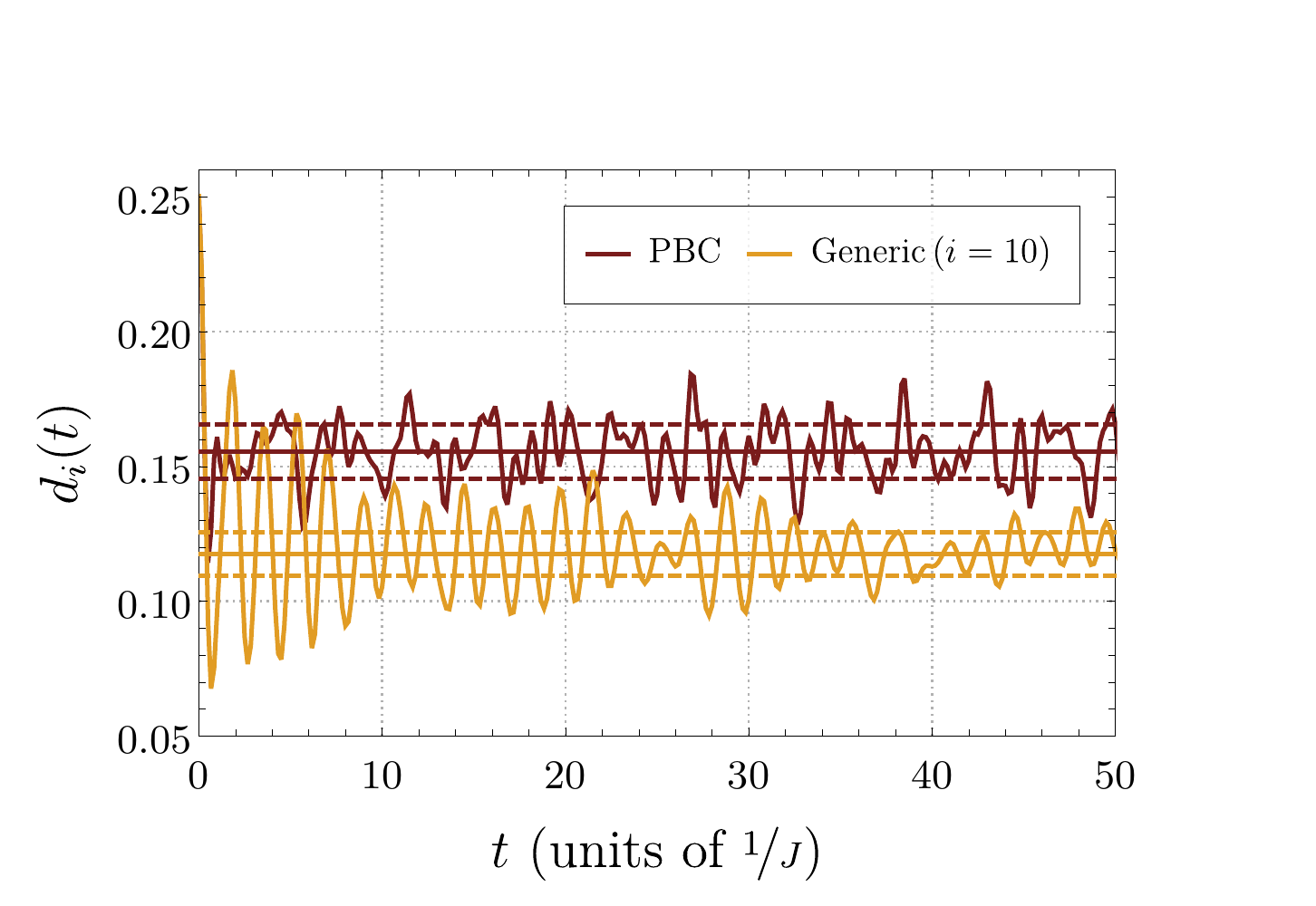}
  \caption{(Color online)\,Time evolution of the double occupancy on the integrable periodic 
	Hubbard chain after interaction quench $U=3J$ and on the non-integrable (Generic) cluster of 
	$N=12$ sites at half-filling. For the non-integrable case, the time evolution of cluster (l), 
	see \Cref{app:finite_clusters}, and its site $i=10$ is depicted and a one percent randomization 
	around the average $U\approx3J$ is chosen. 
	Solid lines denote the average values \eqref{eq:norm_time}, dashed lines the average plus and
	minus the standard deviation $\sigma_i$, both calculated at $\tau=\num{.6}$.}
  \label{img:tpq_cet_thermalization_cet_time_N_12_pbc_vs_generic}
\end{figure}

In the  highly excited state ensuing from the quench 
we examine the tendency of the finite clusters to 
equilibrate by simulating the time-dynamics of proper local observables 
which are measurable in the subsystem $S$. 
To study this phenomenon in detail we consider two types of clusters:
(i) integrable ones with periodic boundary conditions (PBC) and a constant ratio 
$\nicefrac{U}{J}$ as well as (ii) generic clusters with an arbitrary topology. 
A complete overview over the used finite-size clusters is given in \Cref{app:finite_clusters}. 
Henceforth, the labels (a), (b), ... (n) ascribed to the individual topologies will be 
used for identifying a particular cluster. In order to avoid any undesired symmetries 
in the generic clusters, we additionally slightly randomize the parameters of the model such as
the hopping strengths $J_{ij}=J_{ji}$ 
 by drawing their values with uniform probability from the respective intervals 
$[J-p\cdot J;\,J+p\cdot J]$ with $p=0.01$. The same applies to the on-site interactions $U_i$ as well: 
they are taken from the interval $[U-p\cdot U;\,U+p\cdot U]$. 
Note that the randomization is deliberately chosen weak in order to avoid any
many-body localization \cite{Nandkishore2015}. The only purpose of randomization 
is to avoid the influence of accidental symmetries.
In the integrable clusters no randomization is performed because it would spoil the integrability. 

As a meaningful local observable which incorporates two-particle interaction 
we choose the double occupancy $d_i=\bzo{{i\uparrow}}\bzo{{i\downarrow}}$. Thus, the subsystem 
$S$ consists of site $i$. For the calculation of the time-dependence we resort to 
CET as given in \eqref{eq:cet_time_evolution_state}.

Results of the time-dependence in the integrable $\mbox{N=12}$ cluster and in 
the non-integrable cluster (l) of the same size are shown in 
\Cref{img:tpq_cet_thermalization_cet_time_N_12_pbc_vs_generic} for half-filling and 
for $U=3J$. We clearly see signs of the expected fluctuations, see Introduction, 
around an average value without a tendency to converge to a constant stationary value. 
Even on longer time scales (not shown here) no constant stationary value is approached.
This is to be attributed to the finite system size. 

Interestingly, there seem to be indeed qualitative differences between
the integrable and the generic cluster. The time series of the integrable
cluster shows fluctuations which are of the same magnitude for all times.
In contrast, the time series of the generic cluster first shows larger fluctuations
which subsequently  diminish to some extent. This observation, however, 
certainly needs to be substantiated further.

Next, we want to determine the long-time averages of the fluctuating quantities.
These values are the best guesses on finite clusters for stationary values after relaxation.
Since at the beginning there are various transient effects, see 
\Cref{img:tpq_cet_thermalization_cet_time_N_12_pbc_vs_generic},  it is not obvious how
the long-time averages can be computed reliably.
We account for this obstacle by introducing an averaging according to
\begin{equation}
	\label{eq:norm_time}
	\overline{d}(\tau):= \frac{1}{t_\mathrm{max}-t_\mathrm{min}}
	\int_{t_\mathrm{min}}^{t_\mathrm{max}}\dd{t} d(t)
\end{equation}
with $\tau:=\nicefrac{t_\mathrm{min}}{t_\mathrm{max}}\in[0;1]$ for fixed values of 
$t_\mathrm{max}$. By tuning $\tau$ and thus the minimum time starting from which the averaging 
is performed we are able to eliminate the influence of initial relaxation effects 
on the dynamics. If not noted otherwise, all calculations are performed up to 
$t_\mathrm{max}=100/J$.

\begin{figure}[htb]
  \includegraphics[trim=10 10 50 50,clip,width=\columnwidth]{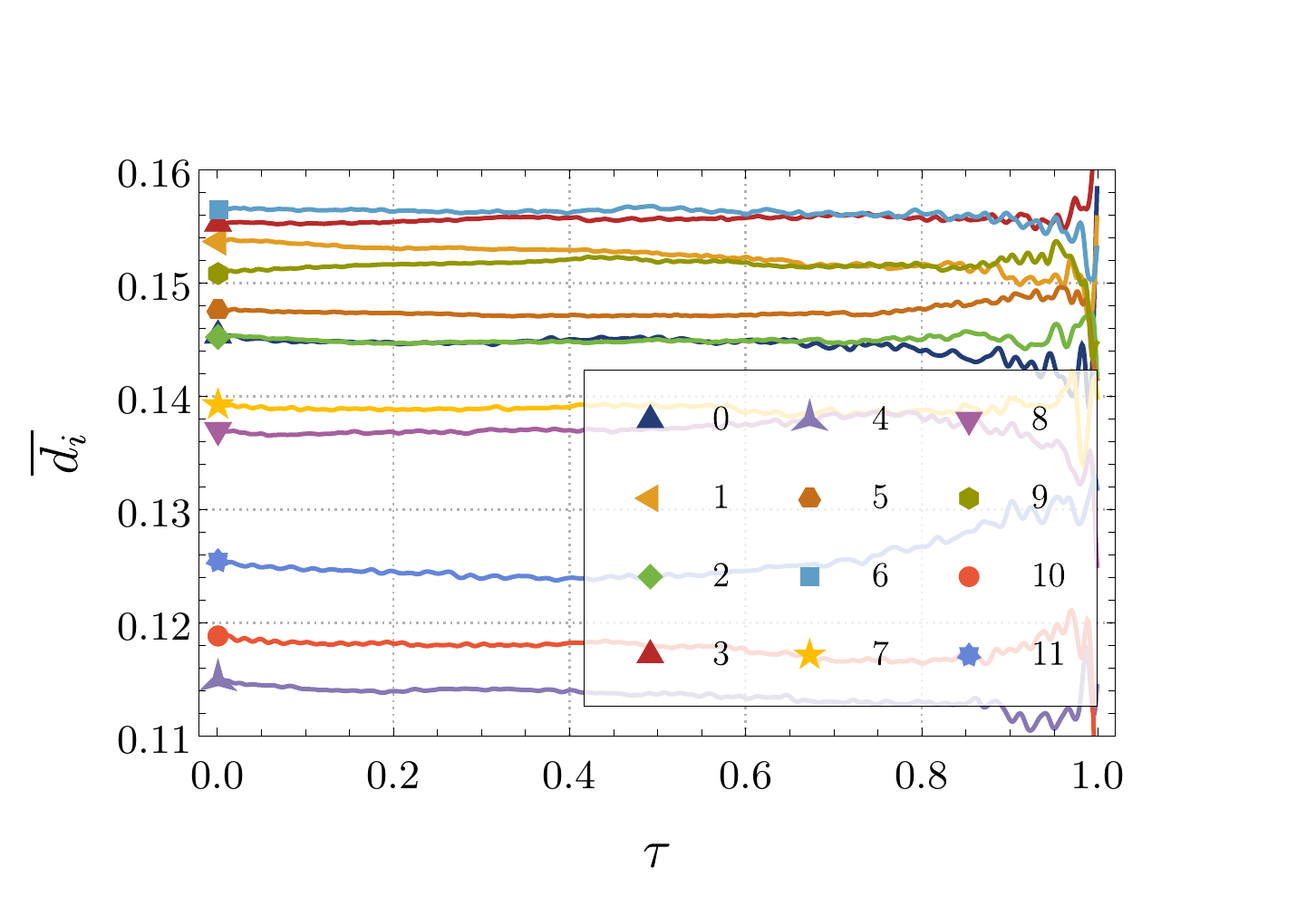}
  \caption{(Color online)\,Averages of the double occupancy of cluster (l) at the sites $i$ determined
	according to 	\eqref{eq:norm_time}. A tendency of the dynamics to converge 
	towards an essentially constant value around $\tau\approx\num{.5}$ to $\tau\approx\num{.6}$ is
	discernible for all sites.}
  \label{img:tpq_cet_thermalization_norm_time_N_12_generic_paper}
\end{figure}

Exemplary results for all sites of the non-integrable, half-filled $N=12$ cluster (l), 
cf. \Cref{app:finite_clusters}, are shown in 
\Cref{img:tpq_cet_thermalization_norm_time_N_12_generic_paper}. As can be seen, some weak initial 
transients are visible up to the range of $\tau\le\num{.2}$. 
During this initial time span we consider the data not fully converged yet, 
cf.\ especially the data for sites $i=6$ or $i=8$. 
After this initial transient, the averaged data converges to an almost constant value.
But choosing $\tau$ too large, i.e., too close to unity, large fluctuations appear.
The reason is that the averaged time span becomes too small so that the fluctuations do 
not cancel sufficiently anymore, cf.\ the range $\tau\gtrapprox\num{.8}$ in 
\Cref{img:tpq_cet_thermalization_norm_time_N_12_generic_paper}. 
In conclusion, avoiding  the initial transient effects as well as the final fluctuations
can be achieved by reading off $\overline{d}_i$ 
for medium values, i.e., around $\tau\approx\num{.5}$ to $\tau\approx\num{.6}$.

In all checked cases of various lattice sizes $N$ and both 
integrable and non-integrable topology the determination of the time-averaged
value according to \eqref{eq:norm_time} is possible since no significant
variations occur in the range of $\tau\approx\num{.5}$ to $\tau\approx\num{.6}$. 
Thus, all following calculations are performed for a constant $\tau=\num{.6}$. 
In this way, we obtain a
suitable approximation of the stationary value of an observable $\overline{A}$ 
as discussed in \Cref{s:introduction}. We refer to these time-averages in the 
study of equilibration and thermalization.

For visual orientation, \Cref{img:tpq_cet_thermalization_cet_time_N_12_pbc_vs_generic} shows 
the long-time averages (solid lines) and the standard deviations around them (dashed lines), 
both calculated at $\tau=0.6$. The initial dynamics differs qualitatively between the two cases 
considered. The generic model shows longer-lasting transients after the quench. Nevertheless, the 
long-time fluctuations show roughly the same amount of spread. This leads to the hypothesis that 
fluctuations show no pronounced dependence on the integrability of the model. We will substantiate 
this conjecture in the following.

The fluctuations present in the dynamics of the system around the time-averaged values 
$\overline{d}_i$ of the double occupancies are quantified by the individual variances 
$\sigma^2_i$. They are a measure for how well the (finite) system stays close to the
time average $\overline{d}_i$. A fully equilibrating system would show vanishing fluctuations 
since it would fulfill $\lim_{t\to\infty}d_i(t)=\overline{d}_i$ so that
$\sigma^2_i=0$ if the latter is determined for long, ideally infinite, time ranges.
Practically, we use \eqref{eq:norm_time} also for the determination of the $\sigma^2_i$. 
We are not aware of analytic \textit{a priori} predictions of the values of $\sigma^2_i$ in the 
physical situation we are considering, namely a highly entangled initial state in
real space. Applying a scheme similar to \eqref{eq:upper_bound_state_distrance} 
for an observable $O$ leads to an upper bound to its variance \cite{Short2010} given by
\begin{equation}
	\label{eq:variance_upper_bound}
	\sigma_O^2 \le \frac{\Delta(O)^2}{4 d_\mathrm{eff}(\omega)} \le 
	\frac{\norm{O}^2}{d_\mathrm{eff}(\omega)}
\end{equation}
with $\norm{O}$ being the largest absolute eigenvalue of the Hermitian operator $O$ and
\begin{equation}
	\label{eq:delta}
	\Delta(O)=2 \min_{c\in\mathbb{C}}\norm{O-c \mathbb{1}}.
\end{equation}
Unfortunately, these upper bounds \eqref{eq:variance_upper_bound} still require the 
cumbersome calculation of the effective 
dimension as main ingredient which can neither be predicted without a complete diagonalization 
nor estimated except for initial product states of system and bath. 
For this reason, our main interest here is to study to which extent
the considered systems equilibrate after their quench.

\begin{figure}[htb]
 {
 \vspace{-.8cm}
 \centering
 {
  \includegraphics[width=1.06\columnwidth,clip]{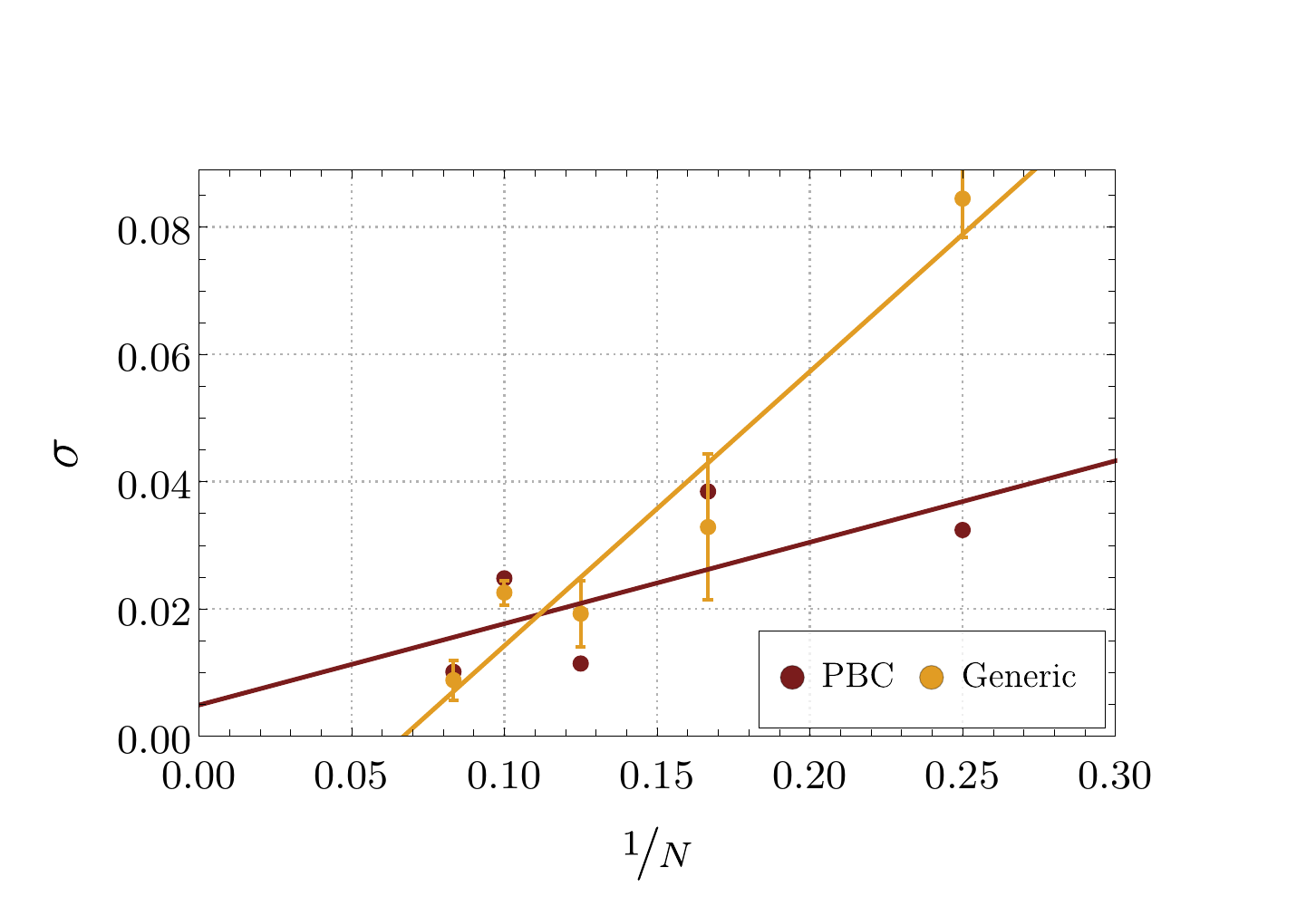}\\
  \vspace{-.7cm}
  \includegraphics[width=1.06\columnwidth,clip]{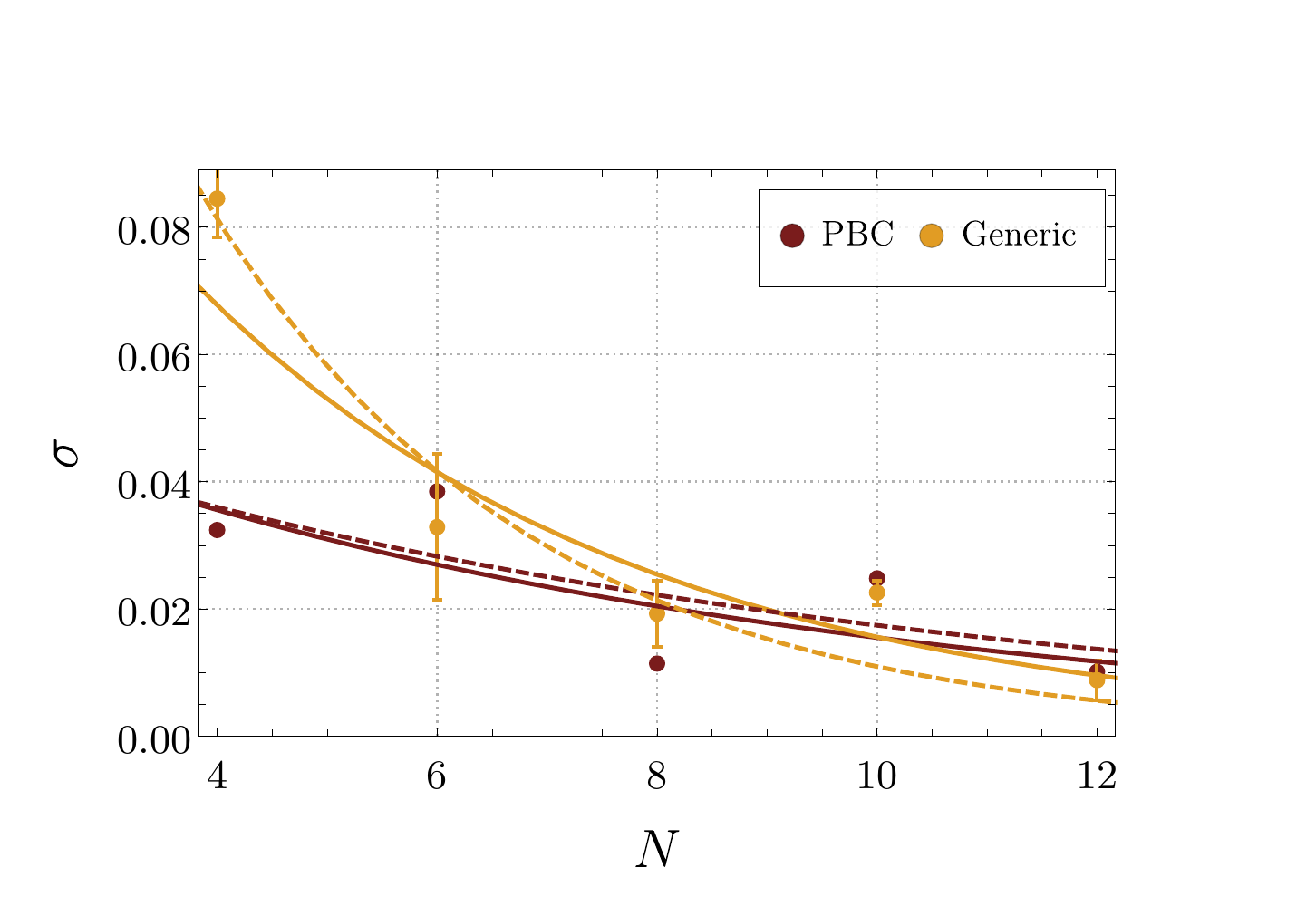}
 }
  \vspace{-.5cm}
  \caption{(Color online)\,Global standard deviations $\sigma$ as derived from \eqref{eq:global-variance} of the 
	double occupancies $d_i(t)$ fluctuating around their individual average values 
 	$\overline{d}_i$. Results for both integrable (PBC) and non-integrable (Generic) clusters are shown. Three different least-square fits to the numerical data are displayed using either $\sigma = a_0+\nicefrac{b_0}{N}$ (upper panel), $\log(\sigma)=\log(a_1)-b_1N$ (lower panel, solid lines) and $\sigma=a_2 \exp(-b_2 N)$ 
	(lower panel, dashed lines). The latter two fits seem to be the same, but this is not the case because 
	the condition of least squares depends on the functional form and leads to differing weights and thus to 
	differing optimum sets $(a_i, b_i)$.
	\label{img:tpq_cet_thermalization_X_error_with_linear_or_exponential_fit_U3}}
	\vspace{-.3cm}
}
\end{figure}

In order not to discuss each site in a cluster separately we define the global variance
\begin{equation}
\label{eq:global-variance}
\sigma^2 = \frac{1}{N}\sum_{i=1}^N\sigma_i^2.
\end{equation}
This quantity provides a good measure for the degree of equilibration.
If it vanishes it indicates equilibration, at least on average. 
\Cref{img:tpq_cet_thermalization_X_error_with_linear_or_exponential_fit_U3} depicts the global 
standard deviation $\sigma$.
For the generic, non-integrable cluster the values for $\sigma^2$ are averaged
additionally over all clusters of the same size $N$, see Appendix 
\ref{app:finite_clusters}, e.g., all generic clusters of $N=12$ sites 
are those labeled by (l)-(n). The plotted error bars indicate the average spread between the maximum and minimum standard deviation for each of the different clusters contributing to each data point for a specific cluster size $N$, i.e., half the error bar amounts to $\nicefrac{1}{2}(\sigma_\mathrm{max}-\sigma_\mathrm{min})$.

The first remarkable observation is that the standard deviations of the integrable and the non-integrable
clusters are very similar for the same cluster size. One could have expected that the fluctuations
in the integrable systems are larger because there is less accessible Hilbert space
due to the large number of conserved quantities. But this does not seem to be the case.
Furthermore, one could think that the similarity of the integrable and non-integrable
fluctuations in \Cref{img:tpq_cet_thermalization_X_error_with_linear_or_exponential_fit_U3} is at odds
with the time series shown in Fig.\ \ref{img:tpq_cet_thermalization_cet_time_N_12_pbc_vs_generic}
where the generic fluctuations are larger briefly after the quench. But for longer
times this is no longer true and it is for these longer times that the quantity $\sigma$
is determined by definition, e.g., the evaluation at $\tau=0.6$ for $t_\text{max}=100/J$
implies that $\sigma$ is computed for the time interval $[60/J,100/J]$. In 
\Cref{img:tpq_cet_thermalization_cet_time_N_12_pbc_vs_generic} the dashed lines and their
mutual distance illustrate that the fluctuations of both systems are comparable in size.

In \Cref{img:tpq_cet_thermalization_X_error_with_linear_or_exponential_fit_U3} 
we tackle the issue to extrapolate the data to the thermodynamic limit. To do so,  we compare two kinds of fits 
with the first one being linear in the inverse lattice size, i.e., $\sigma = a+\nicefrac{b}{N}$ (upper panel) 
and the second one being exponential in the lattice size, i.e., $\sigma=a \exp\left(-b N\right)$
 (lower panel). The exponential fit is carried out in two ways of least-square fits:
(i) $\sigma$ is fitted with $a \exp\left(-b N\right)$, (ii) $\ln(\sigma)$ is fitted
with $\ln(a) - bN$. The difference between both seemingly equal approaches lies in the
least squares which are computed for $\sigma$ or $\ln(\sigma)$ implying different weights.
The first procedure keeps the fit close to the data points
at larger values of $\sigma$ while the second procedure focuses on the data points at
smaller values.

We find that our data is consistent with the exponential scaling predicted \cite{Wilming2019}.
But the numerical data does not provide compelling evidence for the exponential scaling either.
Thus, further study on this issue is certainly called for.
However, both data sets and all fits regardless of the implied form of scaling 
indicate a vanishing global variance for $N\to\infty$. So this provides numerical evidence that equilibration
takes place for systems of increasing system size. Equilibration appears to
be the generic scenario independent of the property of integrability. 
This leads us to conclude that equilibration is an even more generic property than currently proven 
as it is neither limited by a highly entangled initial state 
nor by constants of motion present in integrable systems. 
These conclusions are corroborated by quenches to stronger interactions, for results
see \Cref{app:additional_results} for $U=6J$. For significantly weaker interaction quenches,
the studied time scales and system sizes are not large enough to allow for 
unambiguous evidence, see \Cref{app:additional_results} for $U=J$.

\begin{figure}[htb]
  \includegraphics[trim=10 10 50 50,clip,width=\columnwidth]{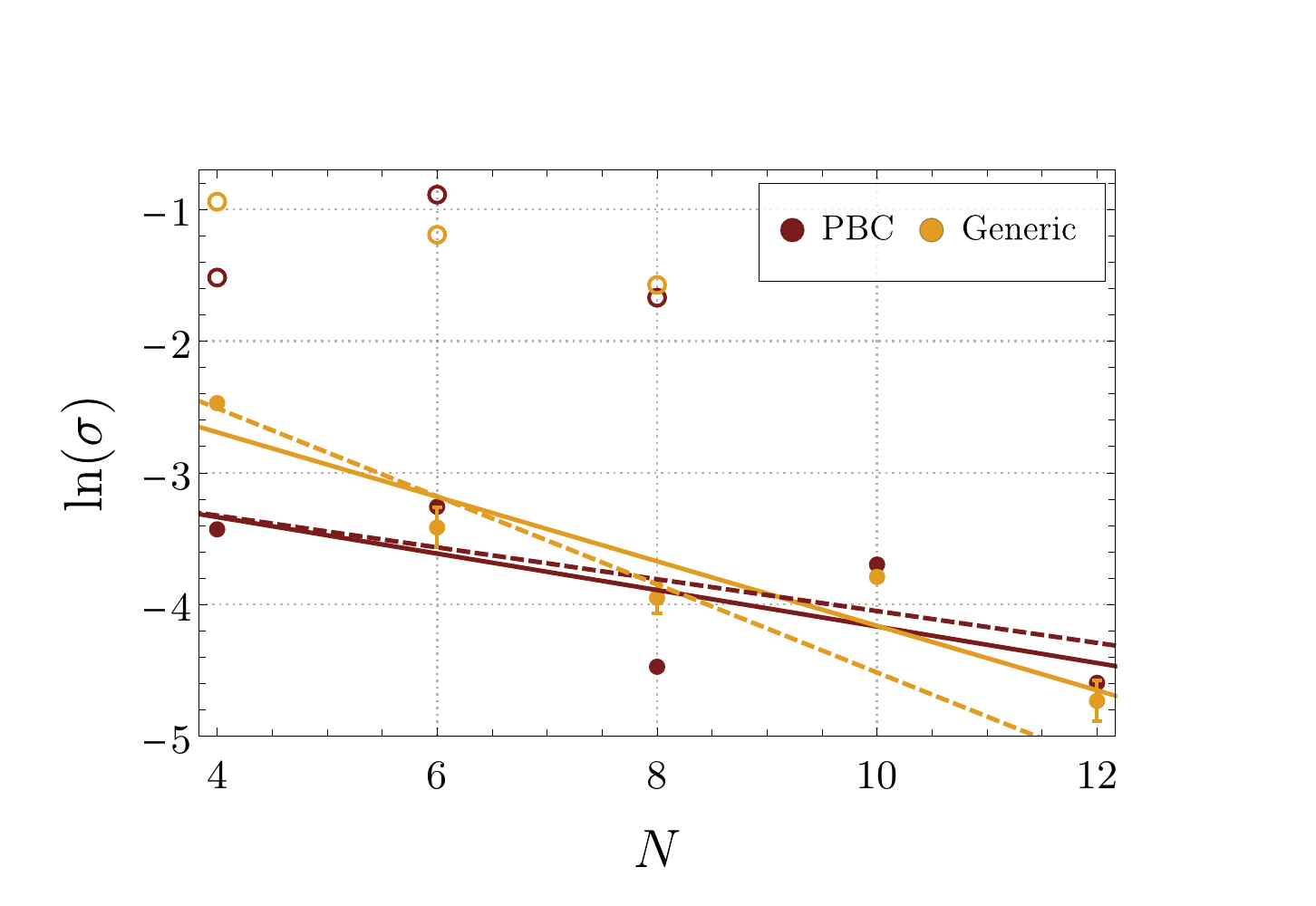}
  \caption{ 
	(Color online)\,Comparison of the actual global standard deviations $\sigma$ (filled symbols) to the upper bounds
	(open symbols) given by \cref{eq:upper_bounds_double_occupancies} on logarithmic scale. The same fits 
	and parameters as in the lower panel of \Cref{img:tpq_cet_thermalization_X_error_with_linear_or_exponential_fit_U3}
	are shown; they appear here as straight lines.}
  \label{img:tpq_cet_thermalization_variance_and_upper_bounds_U3}
\end{figure}

For system sizes that are accessible to complete exact diagonalization we additionally determine the effective dimensions and the respective upper bounds \eqref{eq:variance_upper_bound} to variance and standard deviation. 
In this context, we compare the tightest upper bound for the double occupancies, i.e., $O=d_i$ and 
$c=\nicefrac{1}{2}$ in \eqref{eq:delta}, leading to the upper bound 
\begin{equation}
	\label{eq:upper_bounds_double_occupancies}
	\sigma_i \le \frac{1}{2} d_\mathrm{eff}(\omega)^{-\frac{1}{2}}.
\end{equation}
The required effective dimension is computed assuming the absence of any degeneracy
so that the following relation holds
\begin{equation}
	\label{eq:effective_dimension_by_initial_state}
	\frac{1}{d_\mathrm{eff}(\omega)} = \sum_{n,j} \left(p_j \abs{\braket{n}{\psi_j}}^2\right)^2.
\end{equation}
Here, the initial state may be given as mixture $\rho(0)=\sum_j p_j \dyad{\psi_j}{\psi_j}$ and $\ket{n}$ denote the eigenstates of $H$. 

The upper bounds are displayed by open symbols in the same color as the time-averaged 
standard deviations. The results and fits to the data are shown in 
\Cref{img:tpq_cet_thermalization_variance_and_upper_bounds_U3} on a logarithmic scale. 
It is evident that the mathematically rigorous upper bounds are not particularly tight
for the actually occurring fluctuations.

\begin{figure}[htb]
  \includegraphics[trim=0 0 0 0,clip,width=\columnwidth]{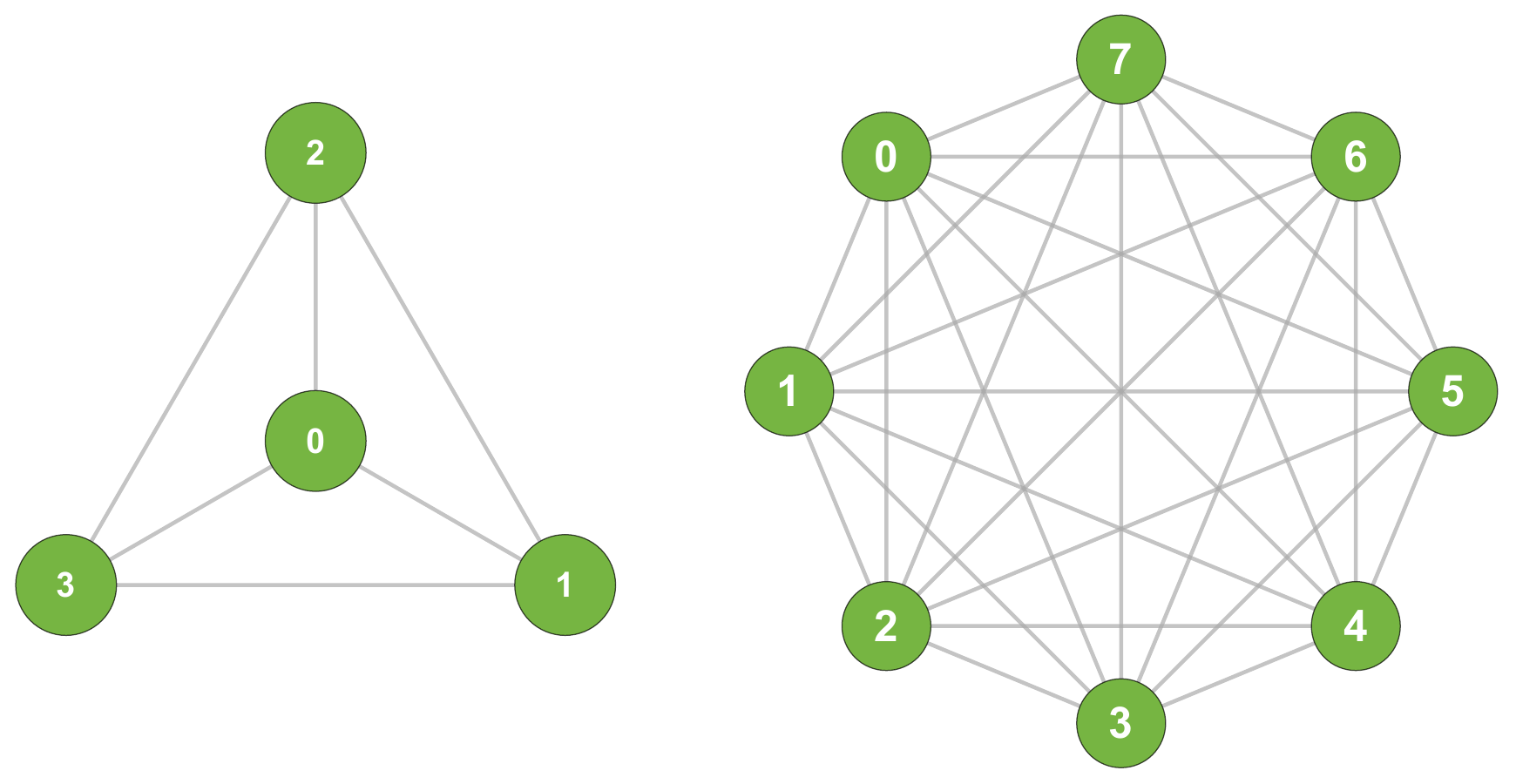}
  \caption{Infinite-range clusters for $N=4$ and $N=8$ denoting clusters with the 
	maximum number of hoppings possible, also called complete graphs. 
	For each of the $N$ sites the coordination number is $z=N-1$ 
	leading to a total of $K=\nicefrac{1}{2} N(N-1)$ hopping links.}
  \label{img:complete_graphs_N_8}
\end{figure}

Discussing fluctuations it is interesting to consider the influence of 
the coordination number $z$. In the clusters considered so far, the typical 
coordination numbers is $z=2$ for the PBC and a mean value of 
$\overline{z}=\num{2.45}$ for the generic clusters. 
Hence these numbers do not vary much. But it is to be expected that systems with
large coordination number display smaller fluctuations. At least in equilibrium, it is common
lore that mean-field approaches work much better in higher dimensions and for larger
coordination numbers because the relevance of the relative fluctuations is lower.
Hence, the same presumption is a plausible working hypothesis out-of-equilibrium.

Here we want to test it for the accessible clusters. Due to the computational limitations
in system size $N$ we choose to consider the limiting case of the maximum value
of the coordination number. It is reached by linking each site with every other site
implying $z=N-1$. The resulting clusters $G_c$
are called infinite-range clusters in physics and complete graphs
in mathematics. In total, they have $K=\nicefrac{1}{2} N(N-1)$ bonds.
 The respective adjacency matrix reads
\begin{equation}
	A(G_c) = J_N - \mathbb{1}_N
\end{equation}
where $J_N$ denotes the $N\times N$ all-ones-matrix and $\mathbb{1}_N$ stands for the identity
 matrix. We subtract the latter one to exclude local terms corresponding to
hops from site $i$ to $i$. We point out that in infinite-range clusters without any randomization 
 the initial Fermi sea state is highly degenerate leading to $\rho^2\ll\rho$. Due to this inherent 
self-averaging the fluctuations in fully symmetric clusters $G_c$ with the same $J$ 
on each bond and the same $U$ at each site are strongly suppressed (not shown). Since this is not 
what we want to study here we again slightly randomize the hoppings $J_{ij}$ and the interactions $U_i$  by 
 \SI{1}{\percent}. This is exactly what we did for the generic clusters
allowing for a study of the direct influence of large coordination numbers
without being distracted by a large number of symmetries.

\begin{figure}[htb]
  \includegraphics[trim=10 10 50 50,clip,width=\columnwidth]{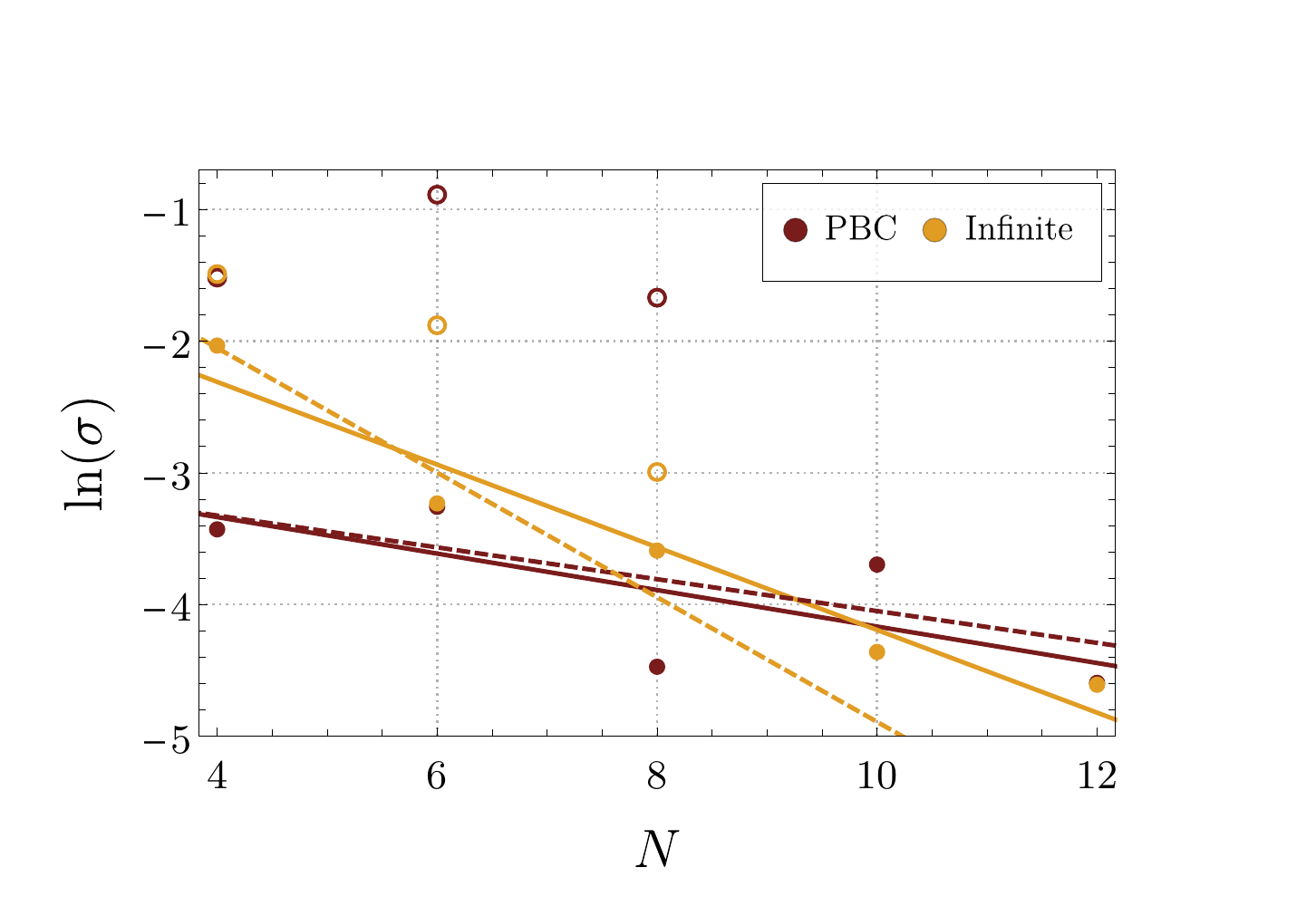}
  \caption{(Color online)\,Global standard deviations $\sigma$ of integrable chains with 
	coordination number $z=2$ and of infinite-range clusters $G_c$ with $z=N-1$ and a one-percent randomization.
	Results are to be compared with \Cref{img:tpq_cet_thermalization_X_error_with_linear_or_exponential_fit_U3}. 
	The amount of fluctuations depends on the number of bonds and decreases
	upon increasing coordination number so that the infinite-range clusters
	display only small fluctuations in the limit $N\to\infty$ relative to the fluctuations in the PBC clusters. 
	The available upper bounds \eqref{eq:upper_bounds_double_occupancies} are inserted using open symbols. Dashed and solid lines are fits, cf.\,\Cref{img:tpq_cet_thermalization_X_error_with_linear_or_exponential_fit_U3}.}
  \label{img:complete_graphs_error_with_fit_U3}
\end{figure}

An example of two infinite-range graphs with $N=4$ and $N=8$, respectively, is given in 
\Cref{img:complete_graphs_N_8}. We use such clusters to compute the time-averaged 
double occupancies $\overline{d}_i$ and subsequently the global standard deviations
 $\sigma$ as before for (non-)integrable models 
in \Cref{img:tpq_cet_thermalization_X_error_with_linear_or_exponential_fit_U3}.
No averaging over various clusters is conducted. 
The results are displayed in \Cref{img:complete_graphs_error_with_fit_U3}
and compared to the ones for integrable chains. Again, we insert the upper bounds for 
$\sigma_i$ determined by \eqref{eq:upper_bounds_double_occupancies} by means of open 
symbols for small systems.  A clear difference of the thermodynamic behavior, i.e., 
for $N\to\infty$, can be noticed.
In PBC systems with a small coordination number the extrapolated fluctuations are noticeably larger
 than in the infinite-range clusters $G_c$.  The standard deviations in the infinite-range clusters 
have a much steeper slope for increasing $N$ 
rendering fluctuations less important for larger  complete graphs than for long PBC chains.
This clearly supports the hypothesis that a larger connectivity favors
smaller fluctuations. Hence, as a rule of thumb we expect that systems with larger
coordination number equilibrate better than those with smaller coordination number.
We stress that this finding does not necessarily imply that the equilibration occurs faster, 
i.e., on a shorter time scale. The issue of time scales is beyond the scope
of the present article since the reliable determination of equilibration time
scales is numerically very challenging.

\subsection{Thermalization}
\label{ss:thermalization}

Here we address the process of thermalization.
In the above \Cref{ss:equilibration} we noted no substantial influence of integrability 
on the degree of equilibration. In both cases of PBC and of the generic clusters the results 
indicated a stationary, equilibrated state in the thermodynamic limit. 
Moreover, the fluctuations due to the finite size of the studied clusters are comparable
for the same system sizes. 

In a next step, it suggests itself to investigate thermalization in
the integrable chains and the generic clusters. To this end, we compare the equilibrated,
time-averaged double occupancies $\overline{d}_i$ with the thermal predictions 
$\expval{d_i}_\mathrm{th}$ where the latter are computed
for the canonical statistical ensemble at the same energy
as the quenched system. Are they equal? In order not to be distracted by
accidental effects at particular sites we define the global deviation
from the thermalized values 
\begin{equation}
	\label{eq:thermal_mean_deviation}
	\Delta_\mathrm{therm} := \frac{1}{N}\sum_{i=1}^N \abs{\overline{d}_i-\expval{d_i}_\mathrm{th}}
\end{equation}
for integrable (PBC) and non-integrable (generic) clusters of size $N$. The thermal predictions 
$\expval{d_i}_\mathrm{th}$ are calculated using TPQ states as described in \Cref{ss:tpq}. 
Since each site $i$ of a given cluster contributes in \eqref{eq:thermal_mean_deviation} 
this definition accommodates for the highly differing individual topologies in a 
systematic way. A system showing perfect thermalization is characterized by a 
vanishing $\Delta_\mathrm{therm} = 0$.
 
\begin{figure}[htb]
  \includegraphics[trim=10 10 50 50,clip,width=\columnwidth]{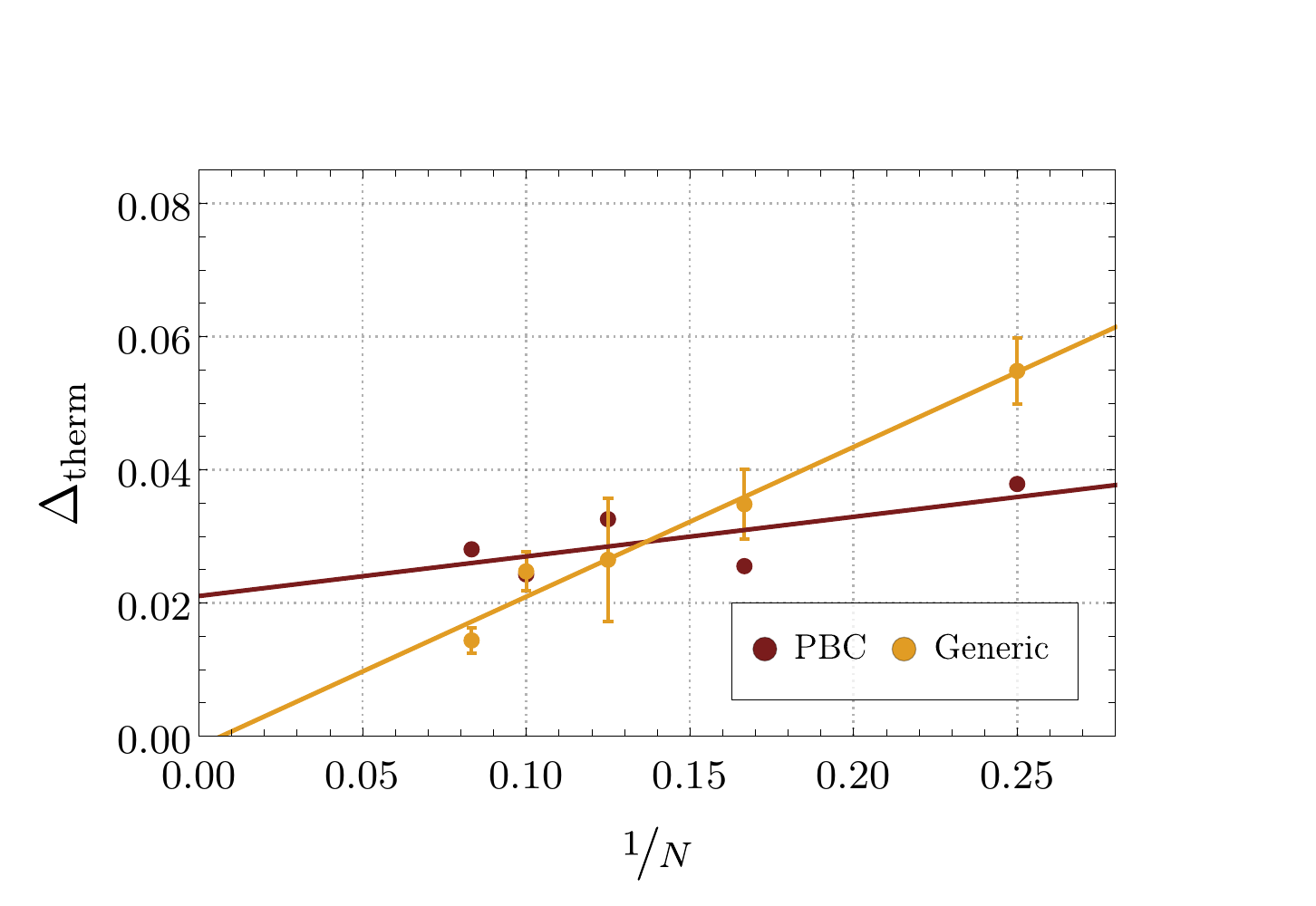}
  \caption{(Color online)\,Global deviation of the time-averages $\overline{d}_i$ from the thermal predictions 
	$\expval{d_i}_\mathrm{th}$ at the effective temperature for $U=3J$. This deviation 
	$\Delta_\mathrm{therm}$ is shown in dependence on the inverse cluster size $N$.
	For the generic clusters the shown values are averaged over various clusters of the same size
	and the error bar indicates the spread within this set of clusters.
	The lines represent linear regressions to the data.}
  \label{img:tpq_cet_thermalization_mean_and_thermal_deviation_with_error}
\end{figure}

Since we are dealing with closed quantum systems the total energy is conserved which allows us 
to determine the effective temperature of the system easily. Knowing the corresponding inverse
temperature $\beta$ is necessary to compute the thermal expectation value of the cluster since it 
defines the statistical density matrix of the canonical ensemble.  The initial state of the system,
 cf.\,\Cref{eq:fermi_sea_state} defines this effective temperature. It has an overall energy 
$E=\mel{\mathrm{FS}}{H}{\mathrm{FS}}$ which translates into an effective inverse temperature
according to \Cref{eq:effective_temperature}.

In \Cref{img:tpq_cet_thermalization_mean_and_thermal_deviation_with_error} the different global
deviations are plotted against the inverse cluster sizes $N$ for the various topologies. 
Error bars again account for the spread of the values between
the differently shaped clusters of same $N$
in the generic, non-integrable cases. In order to analyze the data, a linear fit 
 $\Delta_\mathrm{therm} = \nicefrac{A}{N}+B$ is performed and included in the plot for 
both data sets. In accordance with previous studies
 \cite{Kinoshita2006,Rigol2007,Barthel2008,Kollar2008,Eckstein2008,Rigol2009a,Tang2018}
and with our expectations, clear trends can be read off. 
The generic, non-integrable clusters display a vanishing deviation $\Delta_\mathrm{therm}$ 
in the limit $N\to\infty$. This is a definite indication that theses clusters
thermalize. In contrast, the integrable chains show only a slight decrease of
the global deviation which is not consistent with a vanishing value for
$N\to\infty$. The persisting finite value of $\Delta_\mathrm{therm}$ even for extrapolated
infinitely large systems is a strong sign for equilibration of the integrable chains towards 
a \emph{non-thermal} state. This must be attributed to the restricted dynamics due to the 
large number of  constants of motion.

Since a perfectly thermalizing system loses all of its knowledge about the initial state $\rho(0)$ 
to the larger bath two borderline cases come to ones mind here. First, it is of interest whether a 
system which is only \emph{weakly} perturbed, i.e., which is quenched to $\nicefrac{U}{J}\lessapprox1$, 
is kept from thermalizing. Does a weak quench allow to retain  memory about $\rho(0)$? 
Second, one can wonder whether  quenches even stronger than $U=3J$ also lead to thermalization. 
We discuss both these questions in \Cref{app:additional_results} for brevity.

\section{Summary}
\label{s:summary}


Using numerically exact methods we computed results for equilibration and thermalization 
of arbitrarily shaped finite-size clusters of the quenched Fermi-Hubbard model.
The chosen initial state is the Fermi sea which is highly entangled in real space.
The double occupancy is the local quantity of which the
non-trivial quantum dynamics is studied after the interaction quenches.

We showed that even for the Fermi sea, which is a quantum state extremely far from
a product state in real space, equilibration towards a stationary state is a 
generic property regardless of topology or integrability in the thermodynamic limit, 
i.e., for infinite system sizes. The fluctuations present in the finite systems are
of comparable magnitude for various topologies and do not show a strong influence
of integrability.

In addition, we studied infinite-range graphs which represent systems with maximum 
coordination number at given system size. It was found that the fluctuations in 
these graphs become significantly smaller for $N\to\infty$ than those in graphs of 
coordination number $z=2$. We stress that in infinite-range graphs the coordination 
number increases with system size $z=N-1$.
This corroborates the expectation that fluctuations are
less important for higher connectivity of the cluster. This paradigm
is well established at equilibrium and the evidence found indicates
that it holds true as well in non-equilibrium.

Concerning thermalization, we confirmed the expectations established in the
literature that it depends decisively on the extent that integrals of motion 
exist. The integrable chains studied do not show thermalization, but stay away
from the thermal canonical ensemble. In contrast, the generic clusters clearly
display thermalization.

Obviously, many issues in the field of equilibration and thermalization still
require intensive investigation. Our data showed that there are clear signs
of transient behavior briefly after the quench before the long-time average values and
variances emerge. For conceptual and practical purposes it is highly desirable
to understand this transient behavior better, for instance by determining
or at least estimating the relevant time scales. Knowledge of the relevant time
scales in turn will help to compute long-time averages and stationary values
with high accuracy. Finally, passing from quenches to more general forms
of time-dependences of closed or open quantum systems represents
 a vast field of research. 

\begin{acknowledgments} 
We gratefully acknowledge financial support of the Konrad Adenauer Foundation (PB) as 
well as the German Science Foundation (DFG) in project space UH 90-13/1 (GSU). 
All calculations were performed on the LiDO3 high performance computing system partially 
funded by the DFG. In the context of LiDO3 we especially thank Sven Buijssen for 
helpful technical support.
\end{acknowledgments}

%
%
%

\begin{widetext}
\begin{appendix} 

\section{Finite clusters}
\label{app:finite_clusters}

Below, all clusters are presented which are used in the study of quenches in the generic 
non-integrable one-band Fermi-Hubbard model at half-filling. 
The $J_{ij}$ and $U_i$ have been chosen randomly in a uniform manner 
within a one-percent range around $J$ and $U$.
The number of sites $N$ increases by two 
row by row starting from $N=4$ and going up to $N=12$.

\begin{longtable}{c@{\hskip .5in}c@{\hskip .5in}c}
  \includegraphics[trim=0 0 0 0,clip,width=.25\textwidth]{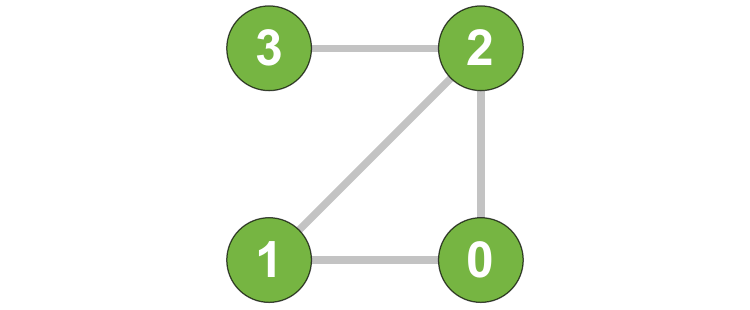} &   
  \includegraphics[trim=0 0 0 0,clip,width=.25\textwidth]{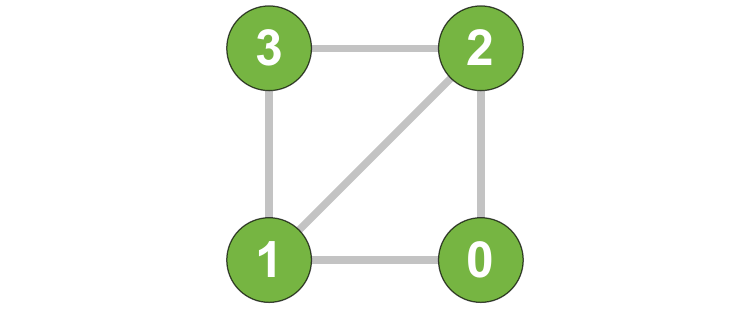} & \\
(a)  & (b)  &  \\[18pt]
 \includegraphics[trim=0 0 0 0,clip,width=.25\textwidth]{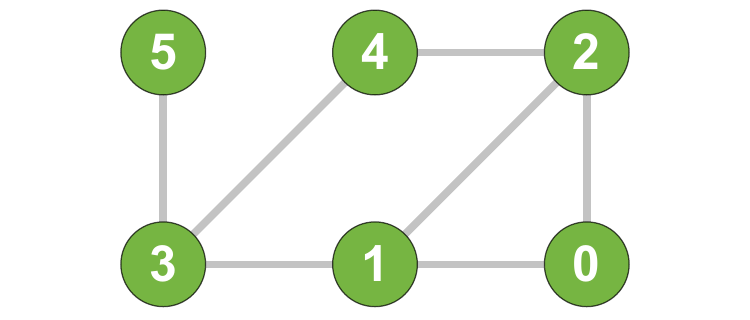} &   
  \includegraphics[trim=0 0 0 0,clip,width=.25\textwidth]{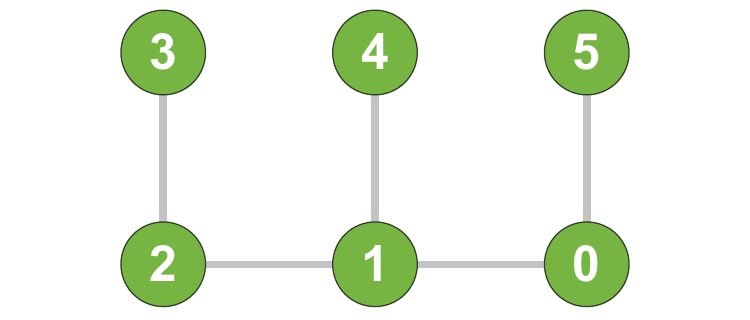} & 
  \includegraphics[trim=0 0 0 0,clip,width=.25\textwidth]{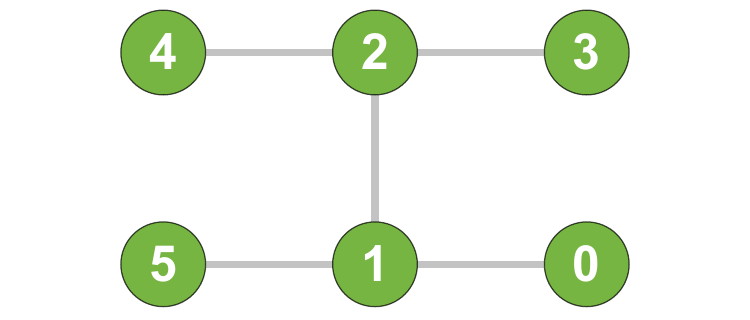} \\
(c)  & (d) & (e) \\[18pt]
\includegraphics[trim=0 0 0 0,clip,width=.25\textwidth]{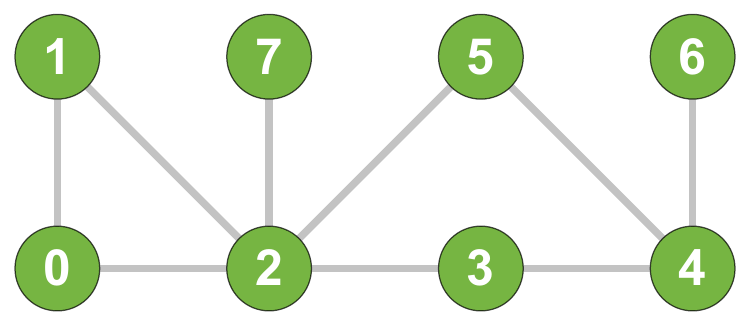} &   
  \includegraphics[trim=0 0 0 0,clip,width=.25\textwidth]{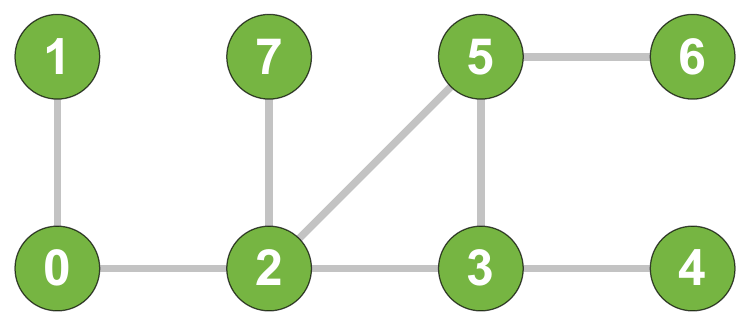} & 
  \includegraphics[trim=0 0 0 0,clip,width=.25\textwidth]{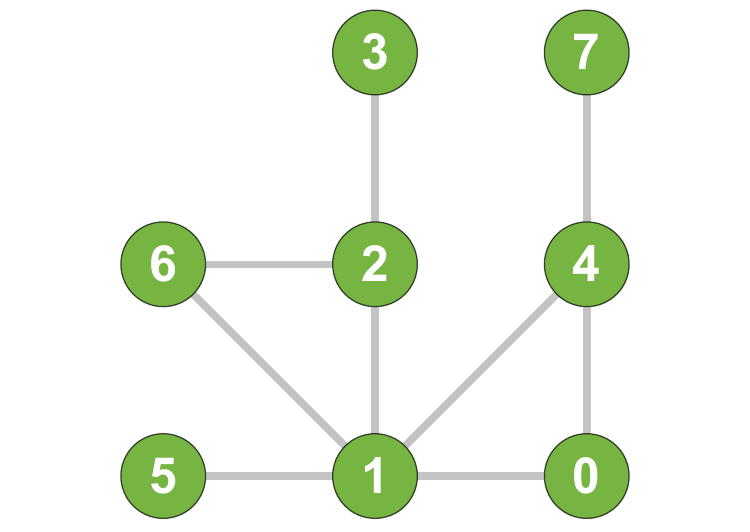} \\
(f)  & (g) & (h) \\[18pt] 
  \includegraphics[trim=0 0 0 0,clip,width=.25\textwidth]{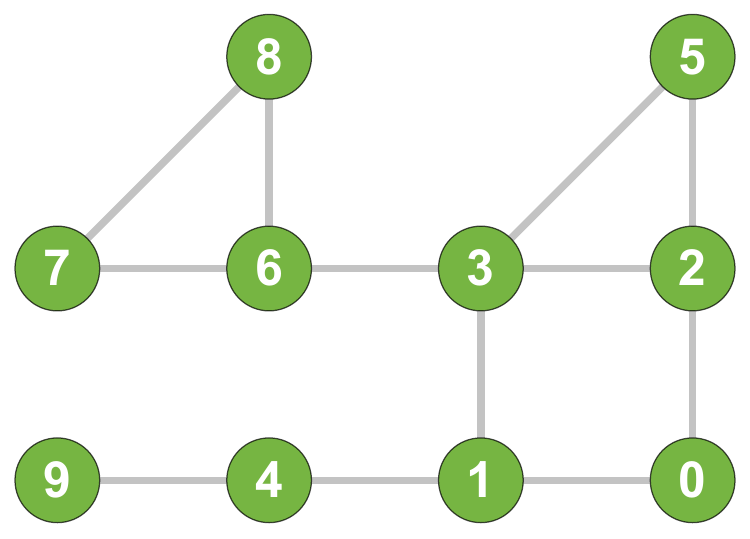} &   
  \includegraphics[trim=0 0 0 0,clip,width=.25\textwidth]{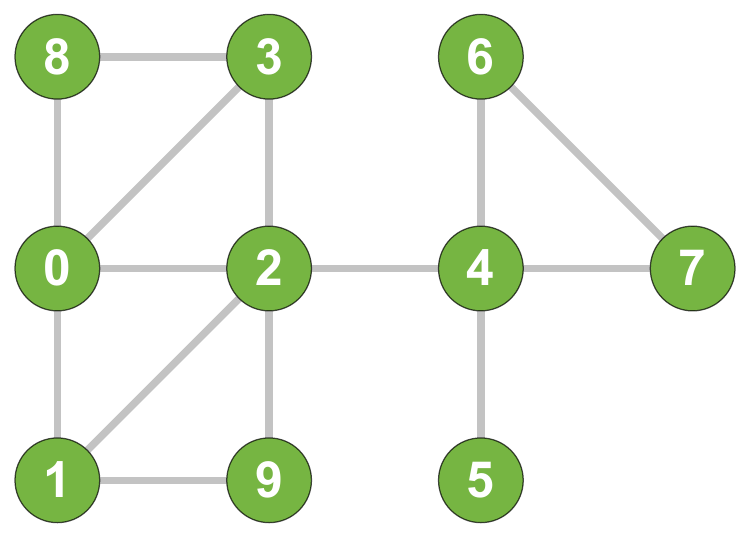} & 
  \includegraphics[trim=0 0 0 0,clip,width=.25\textwidth]{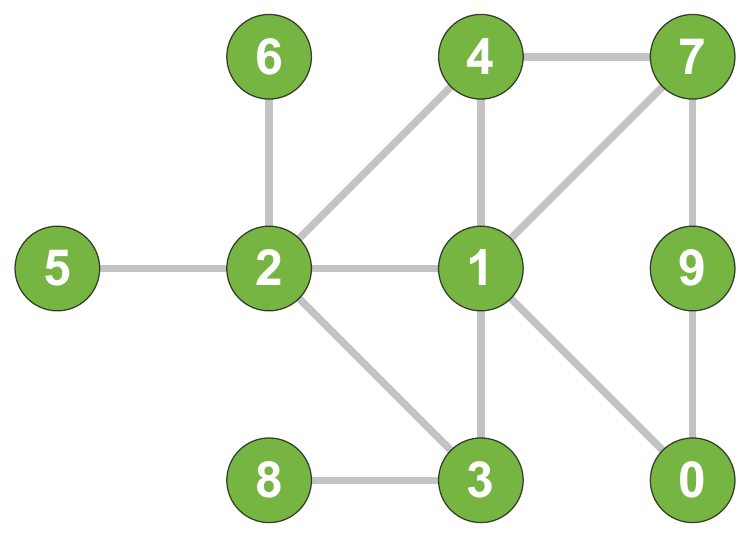} \\
(i)  & (j) & (k) \\[18pt]
  \includegraphics[trim=0 0 0 0,clip,width=.25\textwidth]{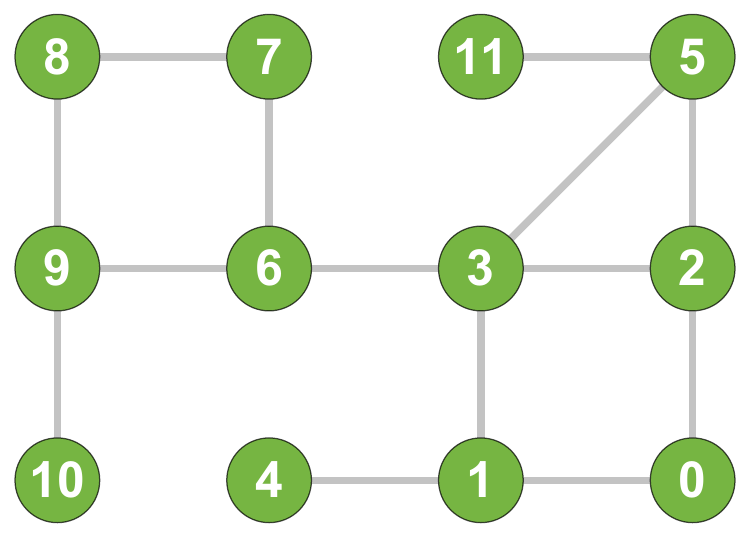} &   
  \includegraphics[trim=0 0 0 0,clip,width=.25\textwidth]{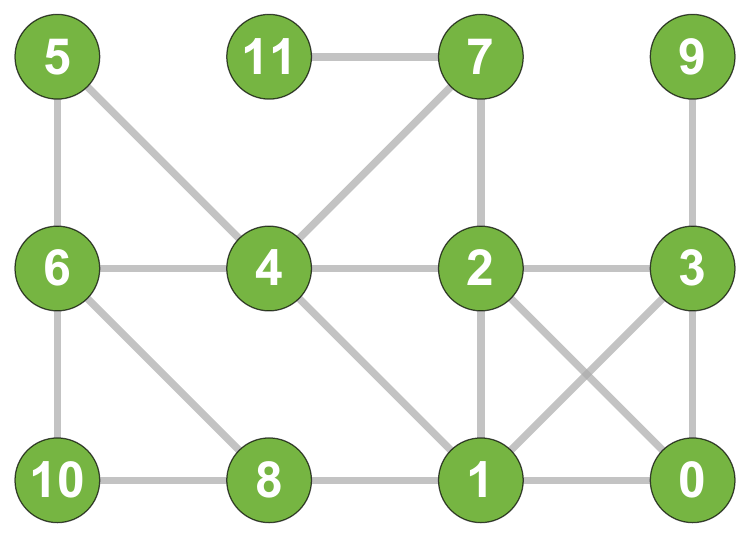} & 
  \includegraphics[trim=0 0 0 0,clip,width=.25\textwidth]{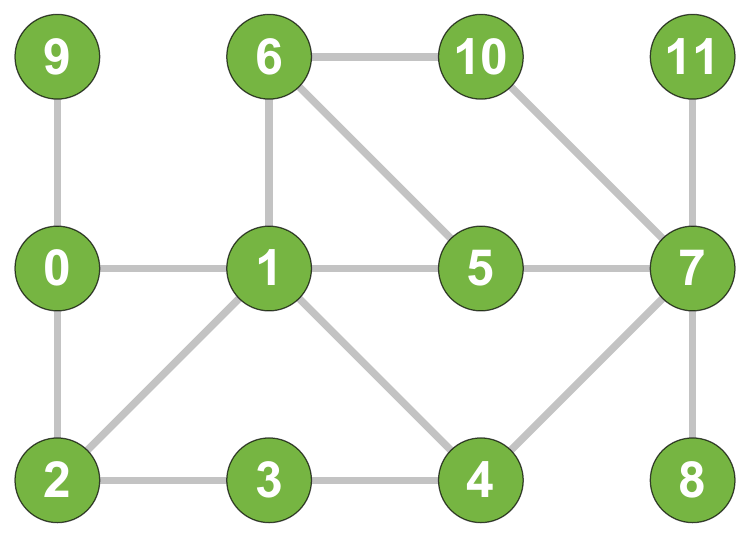} \\
 (l)  & (m) & (n)
\end{longtable}

\section{Additional results for $U=1J$ and $U=6J$}
\label{app:additional_results}

Additionally to the results for sudden interaction quenches to $U=3J$ provided in the main text,
 we simulated quenches to both $U=1J$ and $U=6J$. We again calculated the global standard deviation
$\sigma$ as well as the global deviation $\Delta_\mathrm{therm}$. The respective results are 
thus comparable to the results shown in 
\Cref{img:tpq_cet_thermalization_X_error_with_linear_or_exponential_fit_U3} and 
\Cref{img:tpq_cet_thermalization_mean_and_thermal_deviation_with_error}, respectively. 
Results for $U=1J$ are shown in \Cref{appfig:tpq_cet_thermalization_X_logarithmic_error_and_upper_bounds_U1} 
as well as in \Cref{appfig:tpq_cet_thermalization_X_deviation_U1}.
The results for $U=6J$ are depicted in 
\Cref{appfig:tpq_cet_thermalization_X_logarithmic_error_and_upper_bounds_U6} as well as in 
\Cref{appfig:tpq_cet_thermalization_X_deviation_U6}. 

Both for $U=1J$ and $U=6J$, we again notice a clear tendency of the global standard deviation
$\sigma$ to decrease exponentially with increasing cluster size $N$,
see \Cref{appfig:tpq_cet_thermalization_X_logarithmic_error_and_upper_bounds_U1}
and \Cref{appfig:tpq_cet_thermalization_X_logarithmic_error_and_upper_bounds_U6}.
This is in full accordance with the results of quenches to $U=3J$
and corroborates our conclusion that this is the generic behavior. 

The situation is slightly different for the thermalization behavior characterized by the 
global deviation $\Delta_\mathrm{therm}$ between actual results and thermal predictions. 
The predictions regarding thermalization with a vanishing $\Delta_\mathrm{therm}\to0$ in
 the thermodynamic limit hold when the quenching strength is reasonably large, cf.\
 \Cref{appfig:tpq_cet_thermalization_X_deviation_U6}. In situations, however, where the quench 
is comparably weak -- which is the case when hopping strength $J$ and interaction $U$ are about 
equal at $U=1J$ -- the system is only weakly perturbed. The amount of energy deposited
in the system is relatively small. It is plausible that the effects induced by the lower 
amount of quenched energy make themselves felt only on larger time scales. Concomitantly,
larger spatial scales are also required. While the computations can be done also
for longer times with reasonable effort, increasing linearly in time, it is extremely 
tedious, if not impossible, to tackle larger systems because they have exponentially larger
Hilbert spaces. 

It is worthwhile to notice that the average spread of $\Delta_\mathrm{therm}$ among the generic clusters 
is much larger for $U=1J$ in \Cref{appfig:tpq_cet_thermalization_X_deviation_U1}
than in the other cases $U=3J$ and $U=6J$. This fact emphasizes the higher influence 
of the varying topology of the generic clusters for a particular system size $N$ 
for weak quenches. We attribute this to the fact that for weak interaction quenches
the kinetic part of the Hamiltonian comprising the hoppings remains important.
It is this part which defines the topology; for the local interaction any set of $N$ sites
behaves the same.

\begin{figure}[htb]
\vspace{.4cm}
  \begin{minipage}[b]{0.49\textwidth}
    \includegraphics[trim=10 10 50 50,clip,width=\columnwidth]{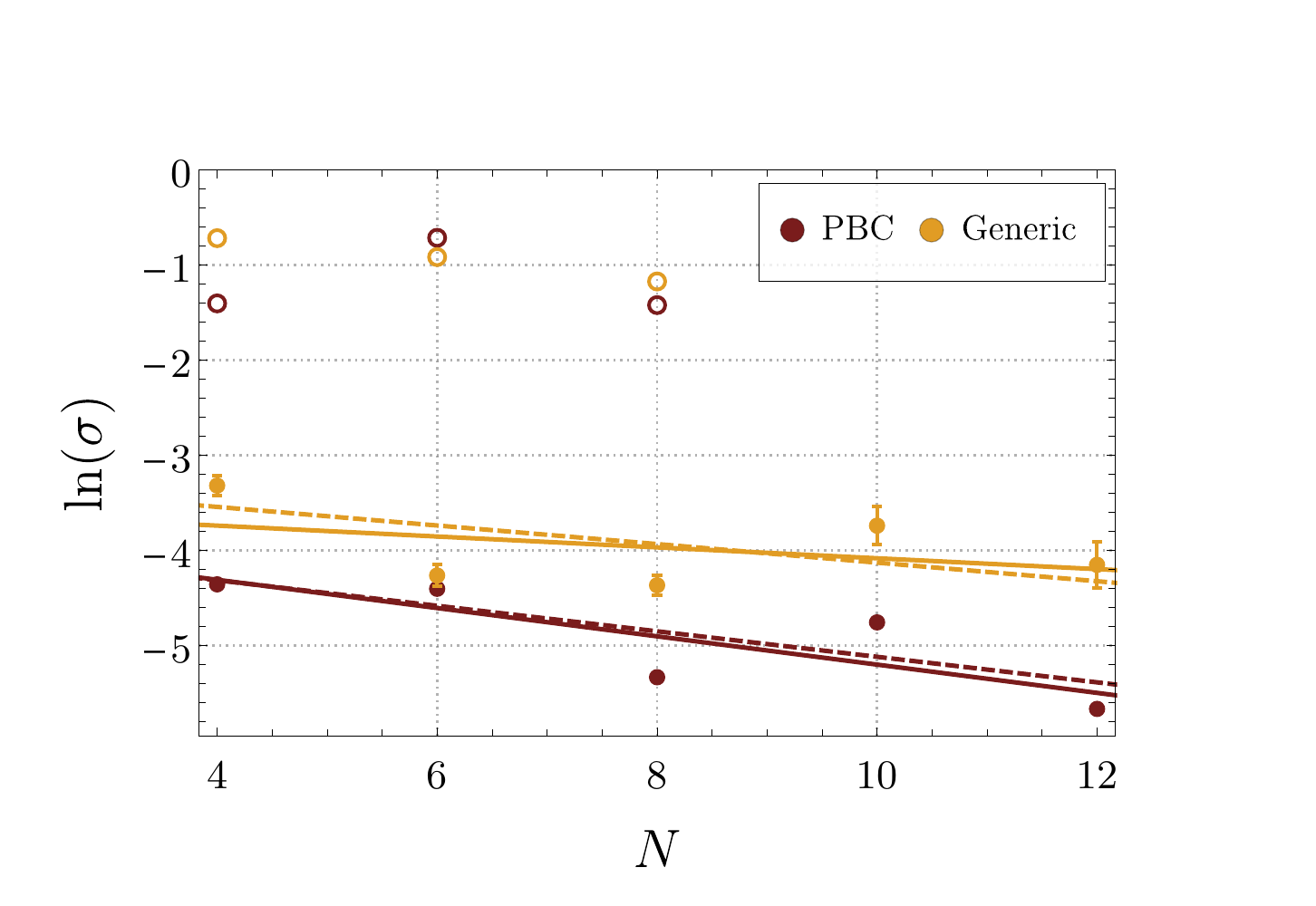}
    \caption{(Color online)\,Results for $U=1J$ showing the global standard deviation $\sigma$ as derived from 
		\eqref{eq:global-variance} of the 	double occupancies $d_i(t)$ fluctuating around their 
		average values 	$\overline{d}_i$ after interaction quenches. In accordance with 
		\Cref{img:tpq_cet_thermalization_X_error_with_linear_or_exponential_fit_U3} 
		fluctuations are becoming exponentially smaller with increasing cluster size $N$. 
		Upper bounds \eqref{eq:upper_bounds_double_occupancies} are computed by complete 
		exact diagonalization and shown using open symbols. 
		The bounds are indeed well above the actual data,  but they are clearly not tight. Solid and dashed lines denote fits, cf.\,\Cref{img:tpq_cet_thermalization_X_error_with_linear_or_exponential_fit_U3}.}
    \label{appfig:tpq_cet_thermalization_X_logarithmic_error_and_upper_bounds_U1}
  \end{minipage}
 \hfill
  \begin{minipage}[b]{0.49\textwidth}
    \includegraphics[trim=5 10 50 50,clip,width=1.03\columnwidth]{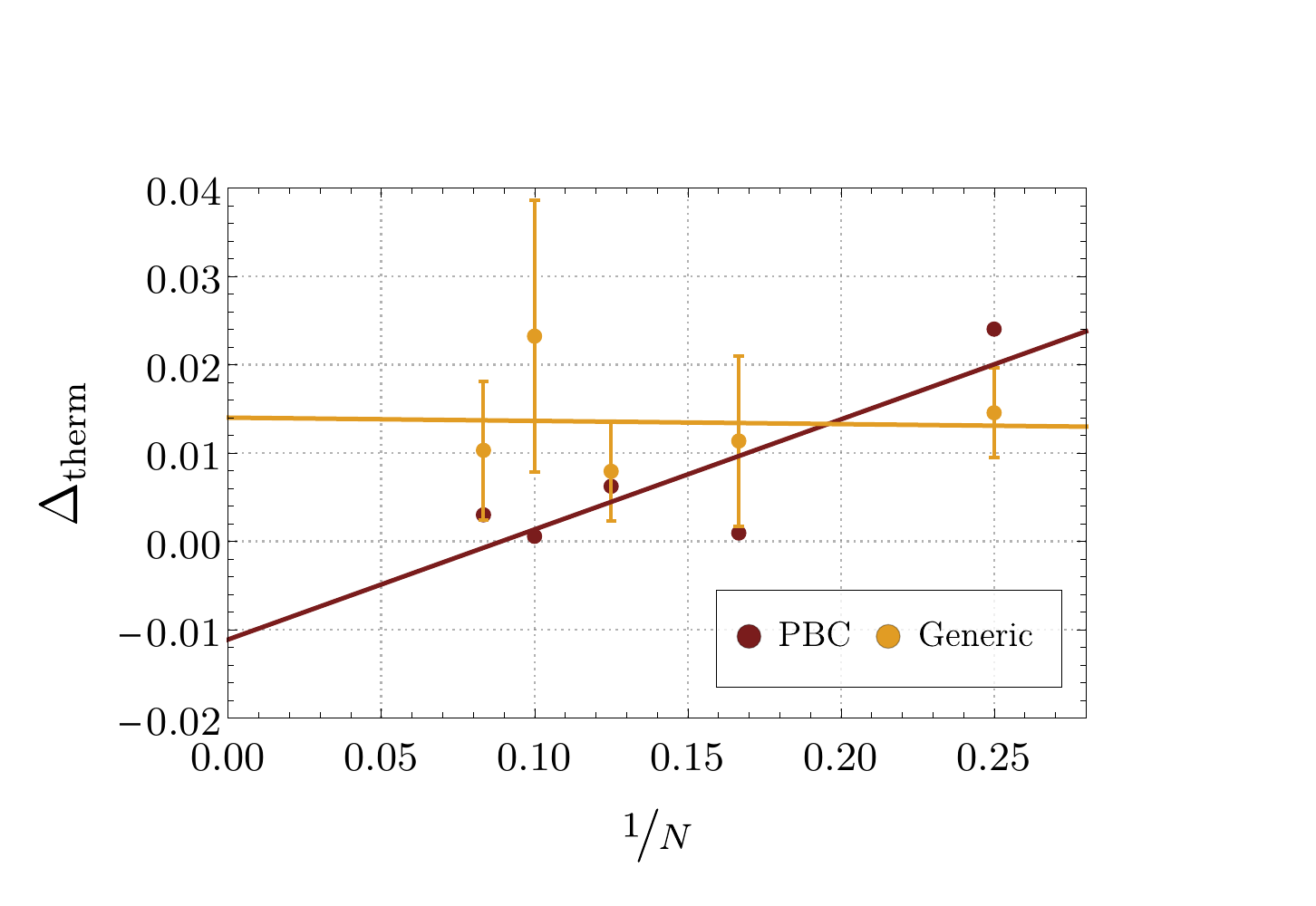}
    \caption{(Color online)\,Results for an interaction quench of the strength $U=1J$ showing the global deviation $\Delta_\mathrm{therm}$ of the 
		time-averages $\overline{d}_i$ from the thermal predictions	$\expval{d_i}_\mathrm{th}$ at 
		the effective temperature of the quench. Integrable (PBC) and non-integrable (Generic) clusters 
		are shown. A pronounced spread between the different generic clusters can be noticed. 
		Since the system is only weakly quenched no clear tendency of thermalization or the absence
		thereof is visible in the data for both the PBC and 
		the generic systems. 
    }
    \label{appfig:tpq_cet_thermalization_X_deviation_U1}
  \end{minipage}
\end{figure}

\begin{figure}[htb]
\vspace{.4cm}
  \begin{minipage}[b]{0.49\textwidth}
    \includegraphics[trim=10 10 50 50,clip,width=\columnwidth]{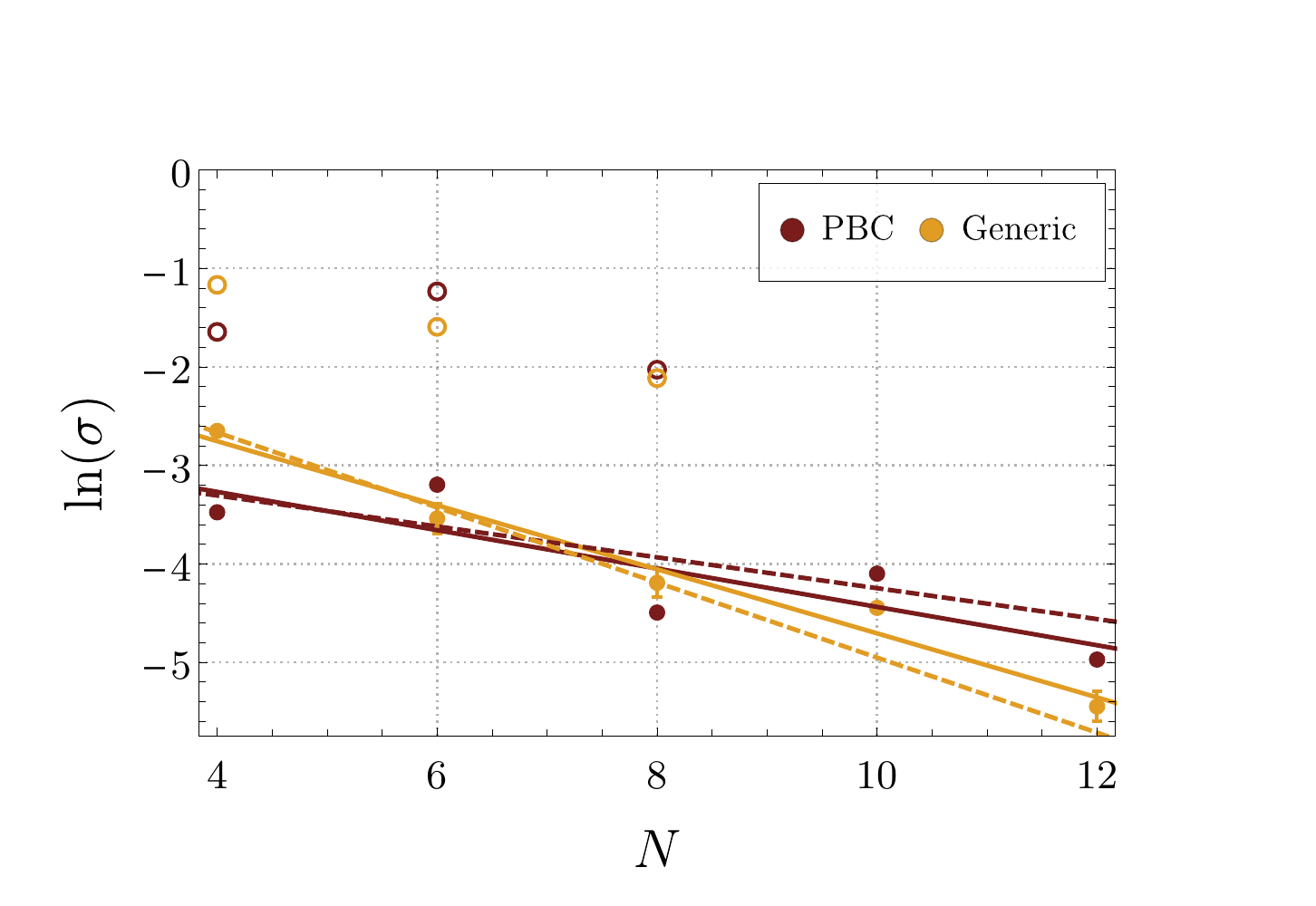}
    \caption{(Color online)\,Global standard deviations $\sigma$ for $U=6J$ as derived from \eqref{eq:global-variance}. 
		The tendency of the fluctuations to decrease exponentially with increasing system size $N$ is obvious. 
		Open symbols show upper bounds \eqref{eq:upper_bounds_double_occupancies}.}
    \label{appfig:tpq_cet_thermalization_X_logarithmic_error_and_upper_bounds_U6}
  \end{minipage}
 \hfill
  \begin{minipage}[b]{0.49\textwidth}
    \includegraphics[trim=10 10 50 50,clip,width=\columnwidth]{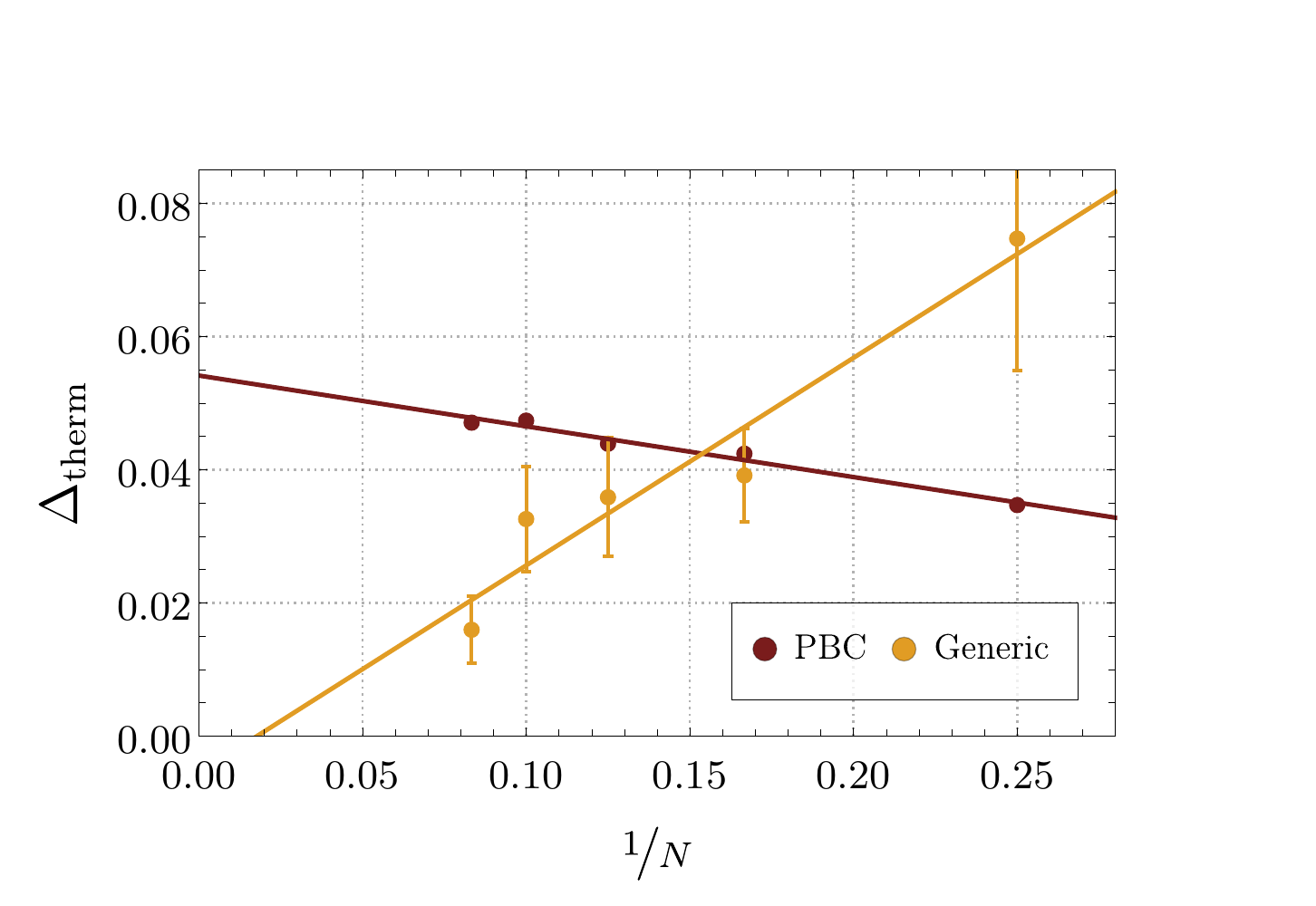}
    \caption{(Color online)\,Global deviation of the time-averages $\overline{d}_i$ from the thermal predictions 
	$\expval{d_i}_\mathrm{th}$ at the effective temperature for $U=6J$. The results agree qualitatively with 
	the ones shown in \Cref{img:tpq_cet_thermalization_mean_and_thermal_deviation_with_error} 
	where a detailed analysis can be found.}
    \label{appfig:tpq_cet_thermalization_X_deviation_U6}
  \end{minipage}
\end{figure}

\end{appendix}
\end{widetext}

\end{document}